\documentclass[12pt, a4paper]{article}
\usepackage{graphicx}
\usepackage{amssymb}
\usepackage{amsmath}
\usepackage{bm}
\usepackage{color}
\usepackage{theorem}
\usepackage{subcaption}
\usepackage{listings}
\usepackage{colortbl}
\usepackage{tabularx}
\usepackage{longtable}
\usepackage[utf8]{inputenc}
\usepackage[T1]{fontenc}
\usepackage{lmodern}
\usepackage[top=30truemm,bottom=30truemm,left=25truemm,right=25truemm]{geometry}

\usepackage[sort&compress,numbers, merge]{natbib}

\definecolor{Orange}{cmyk}{0,0.61,0.87,0}
\definecolor{JungleGreen}{cmyk}{0.99,0,0.52,0}
\definecolor{OliveGreen}{cmyk}{0.64,0,0.95,0.40}
\definecolor{Brown}{cmyk}{0,0.81,1,0.60}
\definecolor{RoyalBlue}{cmyk}{0.71,0.53,0,0.12}
\definecolor{Gray}{cmyk}{0,0,0,0.40}
\definecolor{LightPink}{cmyk}{0.0,0.25,0,0}
\definecolor{LLightPink}{cmyk}{0.0,0.10,0,0}
\definecolor{LightBlue}{cmyk}{0.25,0,0,0}
\definecolor{LightGray}{cmyk}{0,0,0,0.2}

\allowdisplaybreaks[1]

\usepackage[colorlinks=true, linkcolor=OliveGreen, citecolor=RoyalBlue,
urlcolor=RoyalBlue]{hyperref}

\renewcommand{\thefootnote}{\fnsymbol{footnote}}

\begin{document}

\begin{titlepage}

  \begin{flushright}

\end{flushright}

\vskip 1.35cm
\begin{center}

{\large
{\bf
Complementary Probes of Light Higgsinos:
Electroweak \\[3pt] Precision Measurements and Dark Matter Direct Detection
}
}

\vskip 1.5cm

Koichi Hamaguchi$^{a,b}$\footnote{
  E-mail address: \href{mailto:hama@hep-th.phys.s.u-tokyo.ac.jp}{\tt hama@hep-th.phys.s.u-tokyo.ac.jp}}, 
Natsumi Nagata$^a$\footnote{
E-mail address: \href{mailto:natsumi@hep-th.phys.s.u-tokyo.ac.jp}{\tt natsumi@hep-th.phys.s.u-tokyo.ac.jp}},
Genta Osaki$^a$\footnote{
  E-mail address: \href{mailto:osaki@hep-th.phys.s.u-tokyo.ac.jp}{\tt osaki@hep-th.phys.s.u-tokyo.ac.jp}}

\vskip 0.8cm

{\it $^a$Department of Physics, University of Tokyo, Bunkyo-ku, Tokyo 113--0033, Japan} \\[2pt]
{\it ${}^b$Kavli Institute for the Physics and Mathematics of the Universe (Kavli IPMU), University of Tokyo, Kashiwa 277--8583, Japan}

\date{\today}

\vskip 1.5cm

\begin{abstract}

  Although higgsinos are well motivated to be light from the viewpoint of naturalness, they remain difficult to detect experimentally because they interact only through electroweak interactions and typically possess a compressed mass spectrum. 
  While higgsino dark matter can be efficiently probed by direct detection experiments when gauginos are relatively light, the sensitivity rapidly deteriorates for heavier gauginos due to the suppression of higgsino-gaugino mixing. In this paper, we investigate the prospects for probing light higgsinos through future electroweak precision measurements. Focusing on scenarios in which charginos and neutralinos are the only light electroweakly interacting superparticles, we evaluate their contributions to the electroweak oblique parameters as well as to the precision observables \(M_W\) and \(\sin^2\theta_{\mathrm{eff}}\). We compare the projected sensitivities of future \(e^+e^-\) colliders with those of dark matter direct detection experiments. We find that future electroweak precision measurements provide a powerful probe of higgsinos with masses \(\lesssim 500~\mathrm{GeV}\), including parameter regions with highly compressed spectra and spin-independent scattering cross sections below the neutrino fog. On the other hand, dark matter direct detection experiments are particularly sensitive to scenarios with larger charged-neutral mass splittings induced by higgsino-gaugino mixing, and can probe higgsino dark matter all the way up to the thermal relic mass of \(\simeq 1~\mathrm{TeV}\). Our results demonstrate the strong complementarity between electroweak precision measurements and dark matter direct detection experiments in exploring light higgsinos and testing supersymmetric scenarios motivated by naturalness.
  
\end{abstract}

\end{center}
\end{titlepage}

\renewcommand{\thefootnote}{\arabic{footnote}}
\setcounter{footnote}{0}

\section{Introduction}

The discovery of the Higgs boson at the LHC~\cite{ATLAS:2012yve, CMS:2012qbp} completed the particle content of the Standard Model (SM) and established the mechanism of electroweak symmetry breaking. Nevertheless, the origin of the electroweak scale remains one of the central open questions in particle physics. In particular, the Higgs boson mass is highly sensitive to ultraviolet physics, and stabilizing the electroweak scale against large radiative corrections generally requires new physics not far above the weak scale. Supersymmetry (SUSY) provides one of the most compelling solutions to this naturalness problem by introducing superpartners that cancel the quadratic sensitivity of the Higgs mass parameter to heavy scales.

Among SUSY particles, higgsinos play a particularly important role for naturalness. In SUSY models, the electroweak scale is closely related to the higgsino mass parameter $\mu$, and avoiding significant fine-tuning typically requires higgsinos to remain relatively light, often around the electroweak scale~\cite{Ellis:1986yg, Barbieri:1987fn, Kitano:2005wc, Kitano:2006gv, Baer:2012up, Baer:2012cf}. Consequently, light higgsinos are a generic prediction of natural SUSY scenarios, even when other SUSY particles are substantially heavier.

Despite their strong theoretical motivation, however, light higgsinos remain difficult to probe experimentally. Since higgsinos carry only electroweak gauge charges, their production cross sections at hadron colliders are relatively small. Furthermore, in many realistic scenarios the charged and neutral higgsino states are highly degenerate in mass, leading to soft visible decay products and reduced missing-energy signatures. As a result, conventional searches at the LHC suffer from limited sensitivity to compressed higgsino spectra. Although dedicated searches based on soft leptons~\cite{ATLAS:2019lng, ATLAS:2021moa, CMS:2021edw, CMS:2025mie}, disappearing tracks~\cite{ATLAS:2022rme, CMS:2023mny}, or displaced tracks~\cite{Fukuda:2019kbp, ATLAS:2024umc, ATLAS:2025lhc, CMS:2026ias} have improved the sensitivity, a substantial region of parameter space motivated by naturalness remains unexplored.

If the lightest SUSY particle (LSP) is a higgsino-like neutralino, it provides a well-motivated dark matter candidate. In this case, dark matter direct detection experiments offer a powerful and complementary probe of the higgsino sector. In the minimal supersymmetric standard model (MSSM), the dominant contributions to the spin-independent scattering cross section arise from Higgs-boson exchange and are induced by higgsino-gaugino mixing. Consequently, when the gaugino masses are relatively light, current and future direct detection experiments can impose stringent constraints on higgsino dark matter~\cite{Nagata:2014wma, Fukuda:2017jmk, Martin:2024pxx, Martin:2024ytt}. On the other hand, as the gaugino masses increase, the higgsino-gaugino mixing becomes suppressed, leading to a rapid decrease in the direct detection cross section. As a result, even future dark matter direct detection experiments may lose sensitivity to light higgsinos.

In this situation, indirect probes based on precision measurements become increasingly important. Electroweak precision observables (EWPOs) are sensitive to virtual effects of new electroweakly interacting particles through radiative corrections to gauge boson propagators~\cite{Harigaya:2015yaa, DiLuzio:2018jwd, Maura:2024zxz}. Future circular $e^+ e^-$ colliders such as the FCC-ee~\cite{FCC:2025lpp} and CEPC~\cite{CEPCStudyGroup:2018ghi} are expected to achieve extremely precise determinations of electroweak observables. Such measurements provide a promising opportunity to probe light electroweak multiplets even when direct production or dark matter detection becomes challenging.

In this paper, we investigate the prospects for probing light higgsinos through electroweak precision measurements\footnote{For previous studies on the effects of higgsinos on EWPOs, see Refs.~\cite{Grifols:1984xs, Barbieri:1989dc, Drees:1990dx, Drees:1991zk, Barbieri:1991qp, Chankowski:1993eu, Maksymyk:1993zm, Pierce:1996zz, Cho:1999km, Heinemeyer:2004gx, Martin:2004id, Marandella:2005wc, Heinemeyer:2006px, Heinemeyer:2007bw, Nagata:2025ycf, Greljo:2025ggc, Baer:2025tge}.} and compare their sensitivity with that of dark matter direct detection experiments. 
In this comparison, the direct-detection reach is evaluated in our main analysis using the standard thermal relic abundance of higgsinos, with the event rate rescaled by their fractional contribution to the observed dark matter density. The case in which higgsinos constitute all of dark matter is discussed in Appendix~\ref{app:nonthermal}.

We show that future electroweak precision measurements provide a powerful probe of higgsinos with masses \(\lesssim 500~\mathrm{GeV}\), including parameter regions with highly compressed spectra and spin-independent scattering cross sections below the neutrino fog. On the other hand, dark matter direct detection experiments are particularly sensitive to scenarios with larger charged-neutral mass splittings induced by higgsino-gaugino mixing, and can probe higgsino dark matter all the way up to the thermal relic mass of \(\simeq 1~\mathrm{TeV}\). Our results demonstrate the strong complementarity between electroweak precision measurements and dark matter direct detection experiments in exploring light higgsinos and testing supersymmetric scenarios motivated by naturalness.

This paper is organized as follows. In Sec.~\ref{sec:model}, we briefly review the higgsino sector in the MSSM and discuss the resulting chargino and neutralino mass spectrum. In Sec.~\ref{sec:ewpo}, we discuss the contributions of higgsinos to EWPOs in terms of the oblique parameters and describe our analysis framework for future electroweak precision measurements. In Sec.~\ref{sec:dd}, we review the calculation of the spin-independent and spin-dependent dark matter-nucleon elastic scattering cross sections relevant for direct detection experiments. In Sec.~\ref{sec:results}, we present our numerical results and discuss the complementarity between electroweak precision measurements and dark matter direct detection experiments in probing light higgsinos. Finally, Sec.~\ref{sec:conclusion} is devoted to our conclusions. Additional formulae for the vacuum polarization functions are collected in Appendix~\ref{app:vacpol}. The non-thermal dark matter scenario is discussed in Appendix~\ref{app:nonthermal}, while the impact of alternative uncertainty assumptions on the higgsino reach is examined in Appendix~\ref{app:theor_unc}.

\section{Higgsino-like states in the MSSM}
\label{sec:model}

The neutralino and chargino sectors in the MSSM consist of the bino, $\widetilde{B}$, the winos, $\widetilde{W}^a$ ($a = 1,2,3$), and the higgsinos,
\begin{equation}
    \widetilde{H}_d = 
    \begin{pmatrix}
        \widetilde{H}_d^0 \\ \widetilde{H}_d^- 
    \end{pmatrix}
    ~, \qquad 
    \widetilde{H}_u = 
    \begin{pmatrix}
        \widetilde{H}_u^+ \\ \widetilde{H}_u^0 
    \end{pmatrix}
    ~.
\end{equation}
After electroweak symmetry breaking, the neutral fields, $\widetilde{B}$, $\widetilde{W}^3$, $\widetilde{H}^0_d$, and $\widetilde{H}^0_u$, mix with each other to form neutralinos, while the charged fields, $\widetilde{W}^\pm$, $\widetilde{H}_d^-$, and $\widetilde{H}_u^+$, mix to form charginos. We denote the soft SUSY-breaking masses of the bino and wino by $M_1$ and $M_2$, respectively, and the higgsino mass parameter by $\mu$.

In the weak eigenstate basis,
\begin{equation}
    \widetilde{\psi}^0=
    \left(
    \widetilde{B},
    \widetilde{W}^3,
    \widetilde{H}_d^0,
    \widetilde{H}_u^0
    \right)
    ~,
\end{equation}
the neutral fermion mass terms are written as
\begin{equation}
    \mathcal{L}_{\mathrm{mass}}^{(N)}
    =
    - \frac{1}{2}
    \left( \widetilde{\psi}^0 \right)^T
    \mathcal{M}_N
    \widetilde{\psi}^0
    + \mathrm{h.c.}
    ~.
\end{equation}
At tree level, the neutralino mass matrix is given by
\begin{equation}
    \mathcal{M}_N
   =
   \begin{pmatrix}
    M_1 & 0 &
    -M_Z\sin\theta_W \cos\beta &
    M_Z\sin\theta_W\sin\beta \\
    0 & M_2 &
    M_Z \cos\theta_W\cos\beta &
    -M_Z \cos\theta_W \sin\beta \\
    -M_Z\sin\theta_W \cos\beta &
    M_Z \cos\theta_W \cos\beta &
    0 &
    -\mu \\
    M_Z \sin\theta_W \sin\beta &
    -M_Z \cos\theta_W \sin\beta &
    -\mu &
    0
   \end{pmatrix}
   ~,
   \label{eq:mneut}
\end{equation}
where $M_Z$ denotes the $Z$-boson mass, $\theta_W$ is the weak mixing angle, and $\tan\beta \equiv \langle H_u^0 \rangle/ \langle H_d^0 \rangle$ is the ratio of the vacuum expectation values (VEVs) of the neutral MSSM Higgs fields. The neutralino mass matrix is diagonalized by a unitary matrix $N$:
\begin{equation}
 N^* \mathcal{M}_N N^\dagger
 =
 \mathrm{diag}
 \left(
 M_{\widetilde{\chi}^0_1},
 M_{\widetilde{\chi}^0_2},
 M_{\widetilde{\chi}^0_3},
 M_{\widetilde{\chi}^0_4}
 \right)
 ~,
\end{equation}
where the eigenvalues satisfy $M_{\widetilde{\chi}^0_i} \geq 0$. The neutralino mass eigenstates are related to the weak eigenstates through
\begin{equation}
    \widetilde{\chi}^0_i
    =
    N_{ij}\,
    \widetilde{\psi}^0_j
    ~.
\end{equation}

For the charged fermions, the mass terms are given by
\begin{equation}
 \mathcal{L}_{\mathrm{mass}}^{(C)}
 =
 -
 \left(
 \widetilde{W}^-,
 \widetilde{H}^-_d
 \right)
 \mathcal{M}_C
 \begin{pmatrix}
 \widetilde{W}^+ \\
 \widetilde{H}^+_u
 \end{pmatrix}
 +
 \mathrm{h.c.}
 ~,
\end{equation}
where the chargino mass matrix is
\begin{equation}
 \mathcal{M}_C=
\begin{pmatrix}
 M_2 &
 \sqrt{2}\,M_W \sin\beta \\
 \sqrt{2}\,M_W \cos\beta &
 \mu
\end{pmatrix}
~,
\label{eq:mchar}
\end{equation}
with $M_W$ denoting the $W$-boson mass. The chargino mass matrix is diagonalized by unitary matrices $U$ and $V$ as
\begin{equation}
 U^*\mathcal{M}_C V^\dagger
 =
 \mathrm{diag}
 \left(
 M_{\widetilde{\chi}^+_1},
 M_{\widetilde{\chi}^+_2}
 \right)
 ~,
\end{equation}
where the chargino mass eigenstates are defined by
\begin{equation}
 \begin{pmatrix}
  \widetilde{\chi}^+_1 \\
  \widetilde{\chi}^+_2
 \end{pmatrix}
=
V
\begin{pmatrix}
 \widetilde{W}^+ \\
 \widetilde{H}^+_u
\end{pmatrix}
~,
\qquad
 \begin{pmatrix}
  \widetilde{\chi}^-_1 \\
  \widetilde{\chi}^-_2
 \end{pmatrix}
=
U
\begin{pmatrix}
 \widetilde{W}^- \\
 \widetilde{H}^-_d
\end{pmatrix}
~.
\end{equation}

In this paper, we focus on the parameter region $|\mu| \ll |M_1|, |M_2|$, where the lighter chargino and two lighter neutralinos are predominantly higgsino-like. In this limit, the masses of these states are approximately degenerate at tree level: $M_{\widetilde{\chi}_1^+} \simeq M_{\widetilde{\chi}_1^0} \simeq
M_{\widetilde{\chi}_2^0} \simeq |\mu|$, up to corrections suppressed by $M_Z^2/M_{1,2}$. This compressed mass spectrum plays an important role in collider phenomenology and significantly reduces the sensitivity of conventional LHC searches. 

In addition to the tree-level contributions, electroweak radiative corrections induce a mass splitting between the charged and neutral higgsino states. This splitting is phenomenologically important because it controls the decay length of the lightest chargino.
At one-loop level, the dominant contribution to the charged-neutral mass splitting from electroweak gauge boson loops is given by~\cite{Thomas:1998wy}
\begin{equation}
  \Delta m_{+}\big|_{\mathrm{rad}}
  =
  \frac{\alpha_2}{4\pi}
  M_{\widetilde{\chi}_1^+}
  \sin^2 \theta_W
  f \biggl(
  \frac{M_Z}{M_{\widetilde{\chi}_1^+}}
  \biggr)
  ~,
  \label{eq:delmplrad}
\end{equation}
where $\alpha_2 \equiv g^2/(4\pi)$ with $g$ being the $\mathrm{SU}(2)_L$ gauge coupling constant. The loop function $f(x)$ is defined as
\begin{align}
  f(x)
  &=
  2 \int_0^1 dt \,
  (1+t)
  \ln \biggl[
  1 +
  \frac{x^2 (1-t)}{t^2}
  \biggr]
  \nonumber \\
  &=
  - x^2
  + x^4 \ln x
  + 4 x
  \biggl(
  1 + \frac{x^2}{2}
  \biggr)
  \sqrt{1 - \frac{x^2}{4}}
  \tan^{-1}
  \biggl(
  \frac{2}{x}
  \sqrt{
  1 - \frac{x^2}{4}
  }
  \biggr)
  ~.
\end{align}
In the limit $x \to 0$, the loop function behaves as $f(x) \simeq 2\pi x -3x^2 + \mathcal{O}(x^3)$ and thus Eq.~\eqref{eq:delmplrad} is approximated by
\begin{align}
  \Delta m_{+}\big|_{\mathrm{rad}}
  \simeq
  \frac{1}{2}
  \alpha_2
  M_Z
  \sin^2 \theta_W
  \biggl(
  1 -
  \frac{3 M_Z}
  {2 \pi M_{\widetilde{\chi}_1^+}}
  \biggr)
  ~.
\end{align}
For $M_{\widetilde{\chi}_1^+} \gg M_Z$, this mass splitting asymptotically approaches $\Delta m_{+}|_{\mathrm{rad}} \simeq 356~\mathrm{MeV}$.

\section{Electroweak precision measurements}
\label{sec:ewpo}

\subsection{Electroweak precision observables}
\label{sec:ewpos}

In the following analysis, we focus on a scenario in which only neutralinos and charginos give sizable contributions to EWPOs, while sfermions and additional Higgs bosons are assumed to be sufficiently heavy and decoupled. Under this assumption, the dominant effects of neutralinos and charginos arise through loop corrections to the vacuum polarization functions of electroweak gauge bosons. Such effects are conveniently parametrized in terms of the electroweak oblique parameters~\cite{Peskin:1990zt, Peskin:1991sw, Maksymyk:1993zm, Barbieri:2004qk}.\footnote{For the normalization convention of the oblique parameters, we follow Ref.~\cite{Wells:2015uba}.}

The oblique parameters relevant for our analysis are defined as
\begin{align}
  \hat{S}
  &\equiv
  -
  \frac{\cos \theta_W}{\sin \theta_W}
  \Pi^{\prime}_{3B} (0)
  ~,
  \\
  \hat{T}
  &\equiv
  \frac{1}{M_W^2}
  \left[
  \Pi_{WW} (0)
  -
  \Pi_{33} (0)
  \right]
  ~,
  \\
  W  &\equiv -\frac{M_W^2}{2}\Pi^{\prime \prime}_{33} (0)~, 
  \label{eq:wdef}\\
  Y  &\equiv -  \frac{M_W^2}{2} \Pi^{\prime \prime}_{BB} (0)  ~,
  \label{eq:ydef}
\end{align}
where the vacuum polarization functions are defined through
\begin{equation}
  \Pi^{\mu\nu}_{VV'} (q)  =  \Pi_{VV'} (q^2)\,  \eta^{\mu\nu}  -  \Delta_{VV'} (q^2)\,q^\mu q^\nu  ~.
  \label{eq:vpdef}
\end{equation}
Here, $i\Pi^{\mu\nu}_{VV'}(q)$ denotes the one-loop contributions of neutralinos and charginos to the electroweak gauge boson two-point functions, with $V,V' = W, B, 3$ corresponding to the charged weak gauge boson $W^\pm$, the $\mathrm{U}(1)_Y$ gauge boson $B$, and the third component of the $\mathrm{SU}(2)_L$ gauge field, respectively. For completeness, we summarize the analytic expressions for the one-loop contributions of neutralinos and charginos to $\Pi_{VV'}(q^2)$ in Appendix~\ref{app:vacpol}.

The oblique parameters can also be interpreted within the framework of the SM effective field theory. At the dimension-six level, they are related to the Wilson coefficients of the effective operators~\cite{Barbieri:2004qk}
\begin{align}
  \mathcal{L}_{\mathrm{eff}}
  &=
  C_{WB}
  \left(
  H^\dagger \sigma^a H
  \right)
  W^a_{\mu\nu}
  B^{\mu\nu}
  +
  \frac{C_T}{2}
  \left(
  H^\dagger
  \overleftrightarrow{D}_\mu
  H
  \right)^2
  -
  \frac{C_{2W}}{2}
  \left(
  D^\mu W^a_{\mu\nu}
  \right)^2
  -
  \frac{C_{2B}}{2}
  \left(
  \partial^\mu B_{\mu\nu}
  \right)^2
  ~,
\end{align}
through the relations
\begin{align}
  \hat{S} &= \frac{\cos \theta_W}{\sin\theta_W} v^2 C_{WB}  ~,\quad  \hat{T} =  v^2 C_T  ~,\quad 
  W =  M_W^2 C_{2W} ~, \quad 
  Y  = M_W^2 C_{2B} ~,
  \label{eq:ob_wil_rel}
\end{align}
where $H$ denotes the SM Higgs doublet, $W^a_{\mu\nu}$ and $B_{\mu\nu}$ are the field strength tensors of the $\mathrm{SU}(2)_L$ and $\mathrm{U}(1)_Y$ gauge fields, respectively,  $\sigma^a$ ($a=1,2,3$) are the Pauli matrices, ${H}^\dagger \overleftrightarrow{D}_\mu H \equiv H^\dagger D_{\mu} H - (D_\mu H)^\dagger H$, and $v \simeq 246~\mathrm{GeV}$ is the Higgs VEV. Additional electroweak oblique parameters introduced in Ref.~\cite{Barbieri:2004qk}, such as $\hat{U}$, $V$, and $X$, are not generated at the dimension-six level in the effective field theory description. Indeed, in the higgsino-like scenario considered in this work, these parameters are found to be numerically suppressed compared with $\hat{S}$, $\hat{T}$, $W$, and $Y$. We therefore neglect them in the following analysis.

Notice that the \(\hat{S}\) and \(\hat{T}\) parameters in Eq.~\eqref{eq:ob_wil_rel} are proportional to \(v^2\).\footnote{By contrast, the factors of \(M_W^2\) appearing in \(W\) and \(Y\) in Eq.~\eqref{eq:ob_wil_rel} originate simply from the normalization convention adopted in Eqs.~\eqref{eq:wdef} and \eqref{eq:ydef}.} In the chargino/neutralino system, insertions of the Higgs VEV arise only through higgsino-gaugino mixing. Consequently, the contributions to \(\hat{S}\) and \(\hat{T}\) become very small once the gauginos are sufficiently heavy~\cite{Nagata:2025ycf}.

In the following analysis, we also evaluate the corrections to EWPOs induced by neutralinos and charginos. As representative observables, we focus on the $W$-boson mass, $M_W$, and the effective weak mixing angle, $\sin^2 \theta_{\mathrm{eff}}$, which characterize precision measurements from $W$-boson production and the $Z$-pole observables, respectively. Unless otherwise stated, the values of the SM input parameters are taken from Ref.~\cite{ParticleDataGroup:2024cfk}.

The shifts in $M_W$ and $\sin^2 \theta_{\mathrm{eff}}$ can be expressed in terms of the oblique parameters introduced above. Keeping only the leading contributions, we obtain~\cite{Nagata:2025ycf}
\begin{align}
    \frac{\Delta M_W^2}{M_W^2} &\simeq - \frac{2 \sin^2 \theta_W}{\cos 2 \theta_W} \hat{S} + \frac{\cos^2 \theta_W}{\cos 2\theta_W} \hat{T} 
    + \frac{\sin^2 \theta_W}{\cos 2 \theta_W} W + \frac{\sin^2 \theta_W}{\cos 2 \theta_W} Y ~, \label{eq:delmw}\\ 
    \frac{\Delta \sin^2 \theta_{\mathrm{eff}}}{\sin^2 \theta_{\mathrm{eff}}} &\simeq \frac{1}{\cos 2 \theta_W} \hat{S} - \frac{\cos^2 \theta_W}{\cos 2 \theta_W} \hat{T} - \frac{\sin^2 \theta_W}{\cos 2 \theta_W} W - \frac{\cos^2 \theta_W}{\cos 2\theta_W} Y ~.\label{eq:delsw}
\end{align}

\subsection{Future sensitivities and theoretical uncertainties}
\label{sec:uncertainty}

Future lepton colliders are expected to significantly improve the precision of electroweak measurements and thereby provide powerful indirect probes of physics beyond the SM. In particular, the proposed circular $e^+ e^-$ colliders, such as the FCC-ee~\cite{FCC:2025lpp} and CEPC~\cite{CEPCStudyGroup:2018ghi}, aim to achieve unprecedented accuracies in EWPOs owing to their extremely large statistics. Linear colliders, such as the ILC~\cite{ILC:2013jhg} and LCF~\cite{LinearCollider:2025lya}, can also attain high precision through the use of polarized beams, which partially compensate for their lower luminosities by enabling precise measurements of chiral observables.

At circular colliders, the effective weak mixing angle $\sin^2 \theta_{\mathrm{eff}}$ is primarily determined through measurements of forward-backward asymmetries at the $Z$ pole. The FCC-ee is expected to achieve the precision~\cite{FCC:2025lpp}
\begin{equation}
  \Delta \sin^2\theta_{\mathrm{eff}}^{\mathrm{FCC\text{-}ee}}
  =
  1.2~(1.2)\times 10^{-6}
  ~,
\end{equation}
where the numbers outside and inside the parentheses correspond to the projected statistical and systematic uncertainties, respectively. A comparable sensitivity is also anticipated at the CEPC~\cite{CEPCStudyGroup:2018ghi}. At linear colliders, $\sin^2 \theta_{\mathrm{eff}}$ can be extracted from the left-right asymmetry associated with the electron couplings to the $Z$ boson, taking advantage of polarized beams. The Giga-$Z$ program at the ILC~\cite{Yokoya:2019rhx} is expected to achieve the sensitivity $\Delta \sin^2 \theta_{\mathrm{eff}} = 4.0\times 10^{-6}$~\cite{ILCInternationalDevelopmentTeam:2022izu}.\footnote{In the most optimistic scenario considered in Ref.~\cite{ILCInternationalDevelopmentTeam:2022izu}, the left-right asymmetry parameter $A_e$ can be measured with both statistical and systematic uncertainties of $\Delta A_e = 1.3\times 10^{-4}$, which corresponds to $\Delta \sin^2 \theta_{\mathrm{eff}} \simeq 1.7\times 10^{-6}$ through the approximate relation $\Delta A_e \simeq 7.9\, \Delta \sin^2 \theta_{\mathrm{eff}}$.} The LCF is expected to provide a similar sensitivity~\cite{LinearColliderVision:2025hlt}, $\Delta \sin^2 \theta_{\mathrm{eff}} = 3.7\times 10^{-6}$.

The $W$-boson mass can be measured with high precision by performing a threshold scan near the $WW$ production threshold or by reconstructing the $W$-boson kinematics at higher collision energies. The FCC-ee is expected to achieve the sensitivity~\cite{FCC:2025lpp}
\begin{equation}
  \Delta M_W^{\mathrm{FCC\text{-}ee}}
  =
  0.18\,(0.16)~\mathrm{MeV}
  ~.
\end{equation}
For the CEPC, ILC~250,\footnote{A dedicated run near the $WW$ production threshold at the ILC is expected to improve the precision of the \(W\)-boson mass measurement, potentially reaching an accuracy of $\simeq 2$~MeV~\cite{Wilson:2016hne}. } and LCF, the projected sensitivities are $\Delta M_W = 0.5$~MeV~\cite{CEPCStudyGroup:2018ghi}, 2.5~MeV~\cite{ILCInternationalDevelopmentTeam:2022izu}, and 0.6 (2.3)~MeV~\cite{LinearColliderVision:2025hlt}, respectively. 

At the level of precision expected at future lepton colliders, theoretical uncertainties become increasingly important. Precise theoretical predictions are required not only for the extraction of EWPOs from experimental data, but also for the interpretation of these observables within the SM and its extensions. In addition, theoretical input is necessary for determining physical quantities from data, such as particle masses and couplings, which are subsequently used in precision calculations. The current theoretical uncertainties are estimated to be~\cite{Freitas:2019bre} 
\begin{align}
  \Delta \sin^2\theta_{\mathrm{eff}}^{\mathrm{th}}
  &\simeq
  4.5\times 10^{-5} ~,
   \nonumber \\
    \Delta M_W^{\mathrm{th}}
    &\simeq
    4.0~\mathrm{MeV}
    ~.
    \label{eq:therrornow}
\end{align}
These uncertainties are significantly larger than the projected experimental sensitivities at the FCC-ee. In order to fully exploit the precision potential of future facilities, the accuracy of theoretical calculations must be improved by roughly one to two orders of magnitude. Achieving such precision will require substantial progress in higher-order perturbative calculations and in the determination of the relevant SM input parameters, making this a highly challenging task.


In the following analysis, we evaluate the sensitivities of future electroweak precision measurements to higgsinos in terms of the oblique parameters introduced in Sec.~\ref{sec:ewpos}. As discussed above, the projected sensitivities of future experiments may depend significantly on the uncertainties in theoretical predictions. To incorporate these effects, we consider the scenario S1 in Ref.~\cite{Greljo:2025ggc}, which corresponds to a conservative estimate of future theoretical progress. This setup includes uncertainties associated with the conversion of experimental measurements into EWPOs, as well as (aggressive) assumptions for the residual uncertainties in SM predictions due to missing higher-order corrections. In this scenario, the projected uncertainties and the correlation matrix for the oblique parameters are given by
\begin{align}
    \begin{pmatrix}
      \sigma_{\hat{S}} & \sigma_{\hat{T}} & \sigma_W & \sigma_Y
    \end{pmatrix}
    = 
    \begin{pmatrix}
      2.42 & 1.29 & 0.48 & 1.55
    \end{pmatrix}
    \times 10^{-5} ~,
    \label{eq:1sigs1}
\end{align}
and
\begin{align}
   \rho = 
   \begin{pmatrix}
     1 & 0.856 & 0.315 & 0.672 \\ 
     0.856 & 1 & 0.154 & 0.352 \\ 
     0.315 & 0.154 & 1 & 0.221 \\ 
     0.672 & 0.352 & 0.221 & 1 
   \end{pmatrix}
   ~,
\end{align}
respectively.
Using these projected uncertainties and the correlation matrix, we construct the following chi-square function:
\begin{equation}
    \chi^2
    =
    \sum_{i,j}
    \Delta \mathcal{O}_i
    \left(
    \sigma^2
    \right)_{ij}^{-1}
    \Delta \mathcal{O}_j
    ~,
    \label{eq:chi2}
\end{equation}
where
\begin{equation}
    \Delta \mathcal{O}_i
    =
    \left(
    \hat{S},\,\hat{T},\,W,\,Y
    \right)
\end{equation}
denotes the predicted shifts in the oblique parameters and $(\sigma^2)_{ij}$ is the covariance matrix constructed from the projected uncertainties and correlation matrix: $(\sigma^2)_{ij} = \sigma_i \rho_{ij} \sigma_j$.

In Appendix~\ref{app:theor_unc}, we also consider alternative benchmark scenarios presented in Ref.~\cite{Greljo:2025ggc}, corresponding to more optimistic assumptions regarding future improvements in theoretical calculations and the control of systematic uncertainties. We compare the resulting sensitivities to higgsinos with those obtained in our baseline analysis. We find that such improvements can extend the higgsino mass reach by up to several hundred GeV. This demonstrates that continued progress in reducing theoretical and systematic uncertainties will be crucial for fully exploiting the discovery potential of future electroweak precision measurements and establishing sensitivity to higgsino-induced effects.


\section{Dark matter direct detection}
\label{sec:dd}

The sensitivities of dark matter direct detection experiments have improved dramatically in recent years~\cite{LZ:2024zvo, XENON:2025vwd}, leading to increasingly stringent constraints on the scattering cross section between dark matter particles and nucleons. In this section, we briefly review the calculation of the elastic scattering cross section of higgsino dark matter with nucleons.

Let us first consider the spin-independent (SI) scattering cross section. The SI scattering cross section of Majorana fermion dark matter with a nucleon \(N\) is given by
\begin{equation}
  \sigma_{\rm SI}^{(N)}
  =
  \frac{4}{\pi}
  \biggl(
  \frac{m_{\rm DM} m_N}{m_{\rm DM} + m_N}
  \biggr)^2
  f_N^2
  ~,
  \label{eq:sigmasi}
\end{equation}
where \(m_{\rm DM}\) and \(m_N\) denote the dark matter and nucleon masses, respectively, and \(f_N\) is the effective SI coupling between dark matter and the nucleon.

For higgsino dark matter, the dominant contribution to the effective SI coupling arises from Higgs-boson exchange,\footnote{
For higgsino dark matter, the contributions from electroweak loop processes are subdominant~\cite{Hisano:2011cs, Hisano:2012wm, Hisano:2015rsa}. We also note that in the following analysis, we assume that the heavy MSSM Higgs bosons and sfermions are sufficiently heavy and effectively decoupled.
}
which yields
\begin{equation}
  \frac{f_N}{m_N}
  =
  \sum_{q=u,d,s} f_q f_{T_q}^{(N)} +\frac{2}{27} \sum_{Q = c, b, t} f_Q  f_{T_G}^{(N)}  ~,
\end{equation}
where 
\begin{equation}
  f_q = \frac{g^2 }{4 M_W M_h^2} \mathrm{Re}\left[ \left( N^*_{12} - N^*_{11} \tan \theta_W \right) \left( N^*_{13} \cos \beta - N^*_{14} \sin \beta \right)  \right]~,
\end{equation}
with \(M_h\) the Higgs-boson mass. Here, $ f_{T_q}^{(N)} \equiv \langle N | m_q \bar{q} q | N \rangle / m_N$ denote the nucleon matrix elements of the quark scalar operators, and $f_{T_G}^{(N)} \equiv 1 - \sum_{q=u,d,s} f_{T_q}^{(N)}$ represents the gluon contribution induced through the trace anomaly~\cite{Shifman:1978zn}. For the numerical values of the nucleon matrix elements \(f_{T_q}^{(N)}\), we use the default values implemented in \texttt{micrOMEGAs}~\cite{Alguero:2023zol}.\footnote{
For a compilation of nucleon matrix elements and a detailed discussion of the associated theoretical uncertainties in dark matter-nucleon scattering cross sections, see Ref.~\cite{Ellis:2018dmb}.
}

In our analysis, we take the parameters \(\mu\), \(M_1\), and \(M_2\) to be real and $M_1,\, M_2$ to be positive. In the case of higgsino-like dark matter, $|\mu| < M_{1,2}$, and for $M_Z \ll M_{1,2} - |\mu|$, the effective coupling $f_q$ is approximately given by
\begin{equation}
  f_q
  \simeq
  -
  \frac{g^2}{8M_h^2}
  \left[
  \frac{1}{M_2 - |\mu|}
  +
  \frac{\tan^2 \theta_W}{M_1 - |\mu|}
  \right]
  \left(
  1 + \mathrm{sgn}(\mu)\sin 2\beta
  \right)
  ~.
\end{equation}
This expression clearly shows that the effective SI coupling is suppressed for large gaugino masses, reflecting the decoupling of higgsino-gaugino mixing. In addition, the coupling becomes suppressed for small \(\tan\beta\) and negative \(\mu\), due to an accidental cancellation in the Higgs coupling to higgsino dark matter.

On the other hand, the spin-dependent (SD) scattering cross section is given by
\begin{equation}
  \sigma_{\mathrm{SD}}^{(N)}
  =
  \frac{12}{\pi}
  \biggl(
  \frac{m_N m_{\mathrm{DM}}}{m_N + m_{\mathrm{DM}}}
  \biggr)^2
  a_N^2
  ~,
  \label{eq:sigsdN}
\end{equation}
where the effective SD dark matter-nucleon coupling \(a_N\) is expressed as
\begin{equation}
  a_N
  =
  \sum_{q=u,d,s}
  d_q \Delta q_N
  ~.
  \label{eq:an}
\end{equation}
Here, \(\Delta q_N\) denote the spin fractions of the nucleon, defined by $2 s_\mu \Delta q_N \equiv \langle N | \bar{q}\gamma_\mu \gamma_5 q | N \rangle$, with \(s_\mu\) being the nucleon spin four-vector. Again, for the numerical values of the spin fractions, we use the default values implemented in \texttt{micrOMEGAs}~\cite{Alguero:2023zol}. For higgsino dark matter, the dominant SD interaction arises from \(Z\)-boson exchange. The effective SD dark matter-quark coupling is given by
\begin{equation}
  d_q = -
  \frac{g^2}{8 M_W^2}
  \left(
  |N_{13}|^2 - |N_{14}|^2
  \right)
  T_q^3
  ~,
\end{equation}
where \(T_q^3\) denotes the third component of the \(\mathrm{SU}(2)_L\) isospin of the quark \(q\). In the higgsino-like limit, $M_Z \ll M_{1,2} - |\mu|$, $|N_{13}| \simeq |N_{14}| \simeq 1/\sqrt{2}$, which leads to a strong suppression of the \(Z\)-boson coupling to higgsino dark matter: $d_q \simeq 0$. As a result, the SD scattering cross section is also highly suppressed in the pure higgsino limit.

\section{Results}
\label{sec:results}

Now we present the results of our numerical analysis. We perform a parameter scan over the following ranges:
\begin{align}
    150~\mathrm{GeV} &< |\mu| < 1000~\mathrm{GeV} ~, \nonumber \\
    1.6 &< \tan\beta < 50~,
    \nonumber \\
    0.5~\mathrm{TeV} &< M_1 < 10~\mathrm{TeV}~,
    \nonumber \\
    \mathrm{sgn}(\mu) &= \pm 1 ~.
\end{align}
In addition, we impose $M_1 > |\mu|$ and the relation $M_2 = 2 M_1$, which is motivated by gaugino mass unification. In this setup, the wino mass satisfies $M_2 \gtrsim 1~\mathrm{TeV}$, allowing the scenario to evade current LHC constraints on electroweakinos~\cite{ATLAS:2024qxh}.

We assume that the LSP is the higgsino-like state and its relic abundance is determined by thermal freezeout and constitutes only a fraction of the observed dark matter density, while the remaining abundance is composed of some other dark matter component~\cite{Bae:2013bva, Bae:2013hma, Bae:2017hlp}. Accordingly, the event rate for dark matter direct detection is rescaled by the factor $\Omega_{\widetilde{H}} h^2/0.12$, where \(\Omega_{\widetilde{H}}\) denotes the relic density of higgsino dark matter. Namely, the effective spin-independent/dependent cross sections relevant for direct detection are given by
\begin{align}
\sigma_{\rm SI/SD,{\rm eff}}^{(N)} = \frac{\Omega_{\widetilde{H}} h^2}{0.12} 
\sigma_{\rm SI/SD}^{(N)}.
\label{eq:sigma_eff}
\end{align} 
We use \texttt{micrOMEGAs v6.3.0}~\cite{Alguero:2023zol} to compute both the thermal relic abundance of higgsino dark matter and the higgsino-nucleon scattering cross sections.

We also note that one may consider an alternative scenario in which the remaining dark matter abundance is supplemented by non-thermally produced higgsinos~\cite{Fujii:2001xp, Gelmini:2006pw, Gelmini:2006pq, Baer:2014eja, Han:2019vxi, Fukuda:2024ddb}. In such a case, the suppression factor associated with the thermal relic density is absent, and consequently the constraints from dark matter direct detection experiments become significantly stronger. We discuss this non-thermal scenario separately in Appendix~\ref{app:nonthermal}. 

\begin{figure}
  \centering
  \subcaptionbox{\label{fig:st} $\hat{S}, \hat{T}$ }{
  \includegraphics[width=0.48\columnwidth]{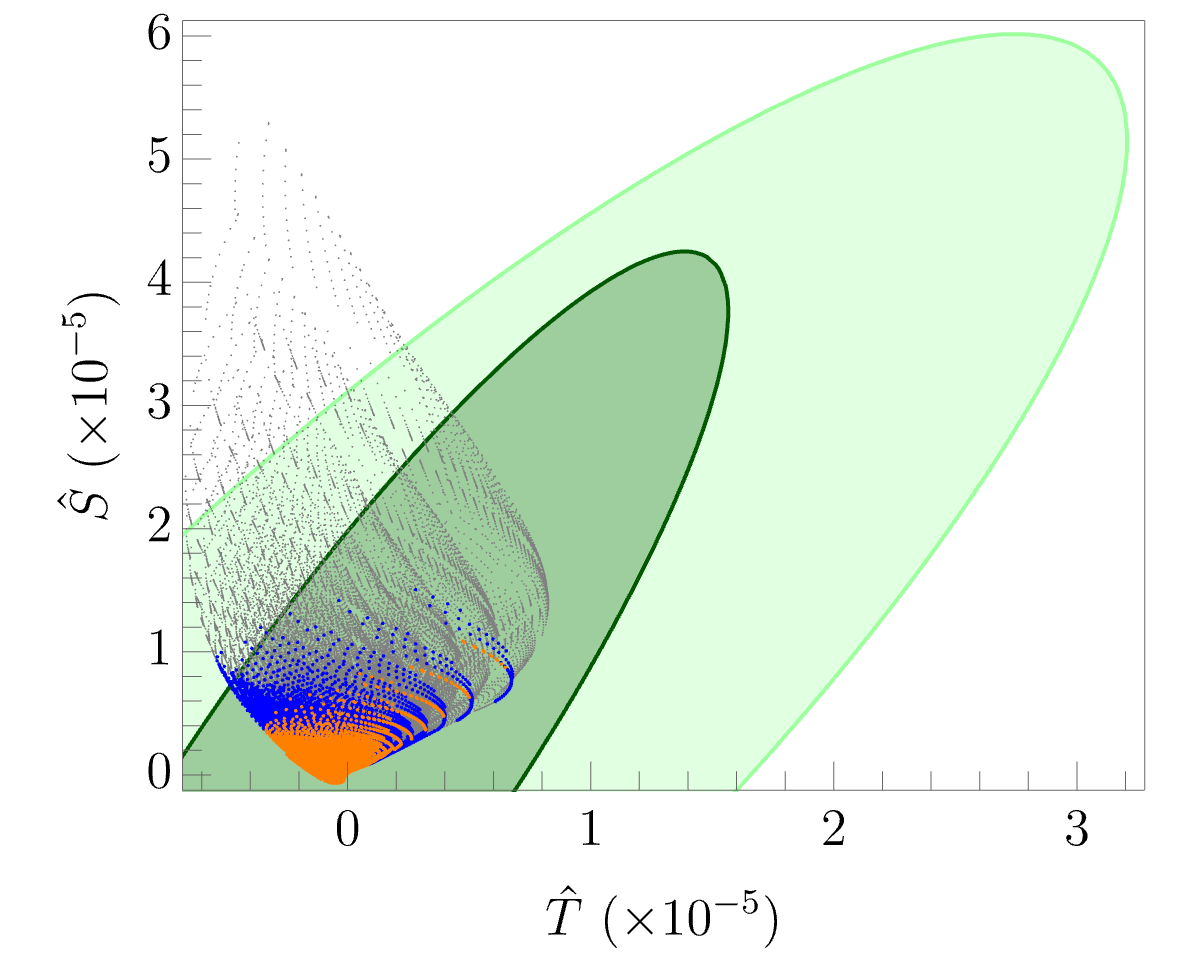}}
  \subcaptionbox{\label{fig:wy} $W, Y$}{
  \includegraphics[width=0.48\columnwidth]{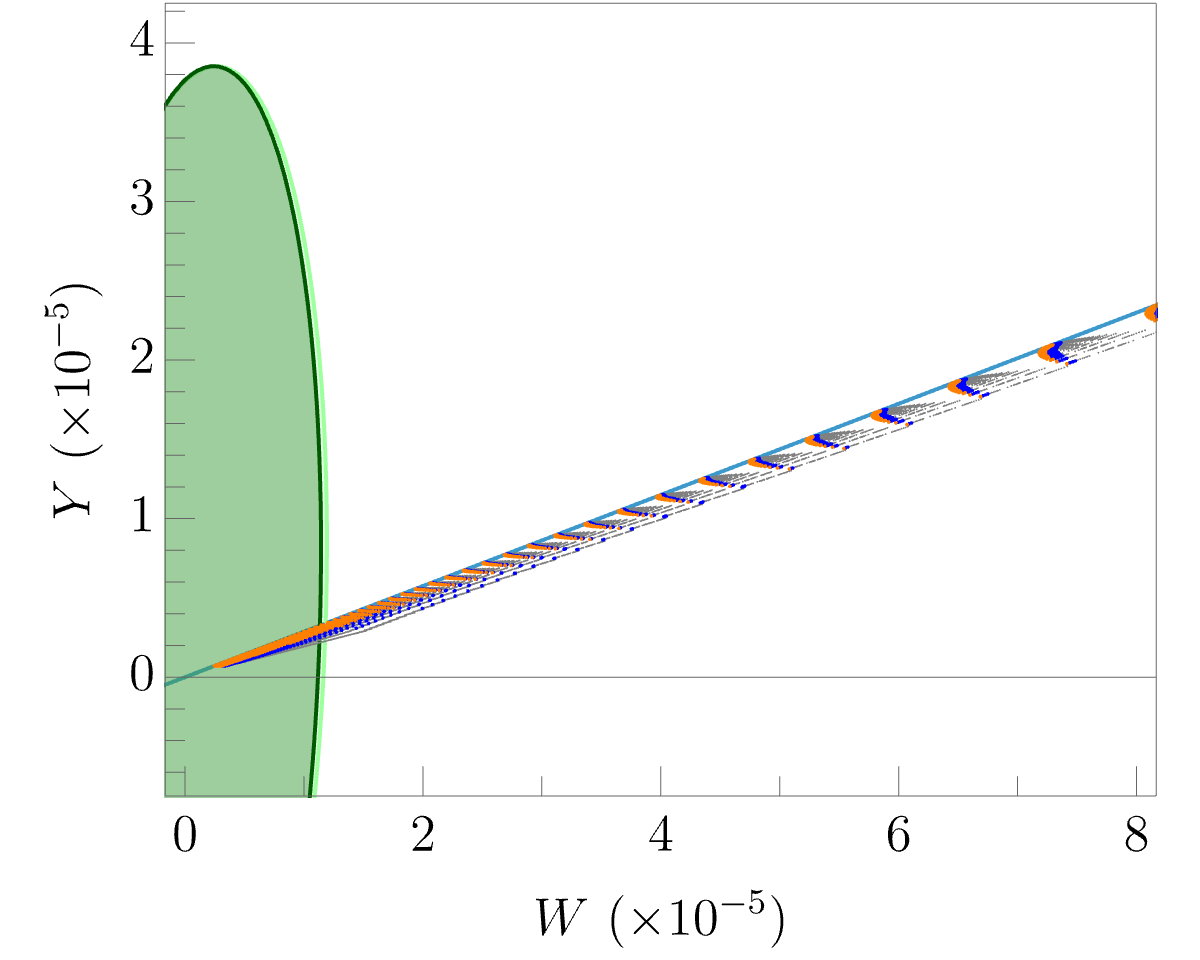}}
\caption{
  Scatter plots of (a) $\hat{S}, \hat{T}$ (b) $W, Y$. The gray points are excluded by the current LZ bound~\cite{LZ:2024zvo}. The orange points correspond to parameter regions lying within the neutrino fog~\cite{OHare:2021utq}, whereas the blue points predict direct detection cross sections above the neutrino fog. The light-green shaded region indicates the projected $2\sigma$ sensitivity corresponding to the S1 scenario introduced in Sec.~\ref{sec:uncertainty}, while the dark-green shaded one corresponds to the pure experimental sensitivity. The blue line in Fig.~\ref{fig:wy} corresponds to the relation in Eq.~\eqref{eq:relyw}.
} 
\label{fig:oblique}
\end{figure}

Figure~\ref{fig:oblique} shows the scatter plots of the oblique parameters obtained in our parameter scan. Figure~\ref{fig:st} displays the distribution in the \(\hat{S}\)--\(\hat{T}\) plane, while Fig.~\ref{fig:wy} shows the distribution in the \(W\)--\(Y\) plane. The gray points are excluded by the current LZ bound~\cite{LZ:2024zvo}. The orange points correspond to parameter regions lying within the neutrino fog~\cite{OHare:2021utq}, whereas the blue points predict direct detection cross sections above the neutrino fog. The light-green region indicates the projected sensitivity corresponding to the S1 scenario introduced in Sec.~\ref{sec:uncertainty}, while the dark-green shaded one corresponds to the pure experimental sensitivity (the S3 scenario given in Appendix~\ref{app:theor_unc}). For these sensitivities, we show the marginalized \(2\sigma\) regions derived from the \(\chi^2\) function in Eq.~\eqref{eq:chi2}.

From Fig.~\ref{fig:st}, we find that, once the current LZ constraints are imposed, the surviving parameter points typically predict only small values of \(\hat{S}\) and \(\hat{T}\). This behavior can be understood from the fact that the direct detection constraints disfavor parameter regions with sizable higgsino-gaugino mixing, which would otherwise induce larger contributions to these oblique parameters. As a result, probing higgsinos through \(\hat{S}\) and \(\hat{T}\) alone becomes challenging even at future electroweak precision experiments. 

On the other hand, Fig.~\ref{fig:wy} shows that the predicted values of \(W\) are sufficiently large to be probed by future electroweak precision measurements over a substantial region of parameter space, including regions that evade current and future direct detection constraints, whereas the \(Y\) parameter typically remains below the projected experimental sensitivities. This demonstrates that the sensitivity of future electroweak precision measurements to higgsinos is predominantly driven by the \(W\) parameter. We also note that the sensitivities to the \(W\) and \(Y\) parameters are largely insensitive to the theoretical uncertainty. The reason is that these parameters are primarily constrained by measurements of fermion pair production at center-of-mass energies above the \(Z\)-pole, where the impact of theoretical uncertainties is expected to be relatively small compared with the anticipated experimental accuracy~\cite{Greljo:2024ytg}. 

Another feature visible in Fig.~\ref{fig:wy} is the strong correlation between the predicted values of the \(W\) and \(Y\) parameters. This behavior can be understood from the approximate expressions for the neutralino and chargino contributions to these parameters in the heavy-gaugino limit, given in Ref.~\cite{Marandella:2005wc}:
\begin{align}
  W &\simeq \frac{g^2}{120 \pi^2} \frac{M_W^2}{\mu^2} ~,
 \qquad 
  Y \simeq \frac{g^2 \tan^2 \theta_W}{120 \pi^2} \frac{M_W^2}{\mu^2} ~.
  \label{eq:wyappr}
\end{align}
These expressions imply the simple relation
\begin{equation}
  Y \simeq \tan^2 \theta_W \, W ~,
  \label{eq:relyw}
\end{equation}
which explains the nearly linear correlation observed in Fig.~\ref{fig:wy}. 

\begin{figure}
  \centering
  \includegraphics[width=0.5\columnwidth]{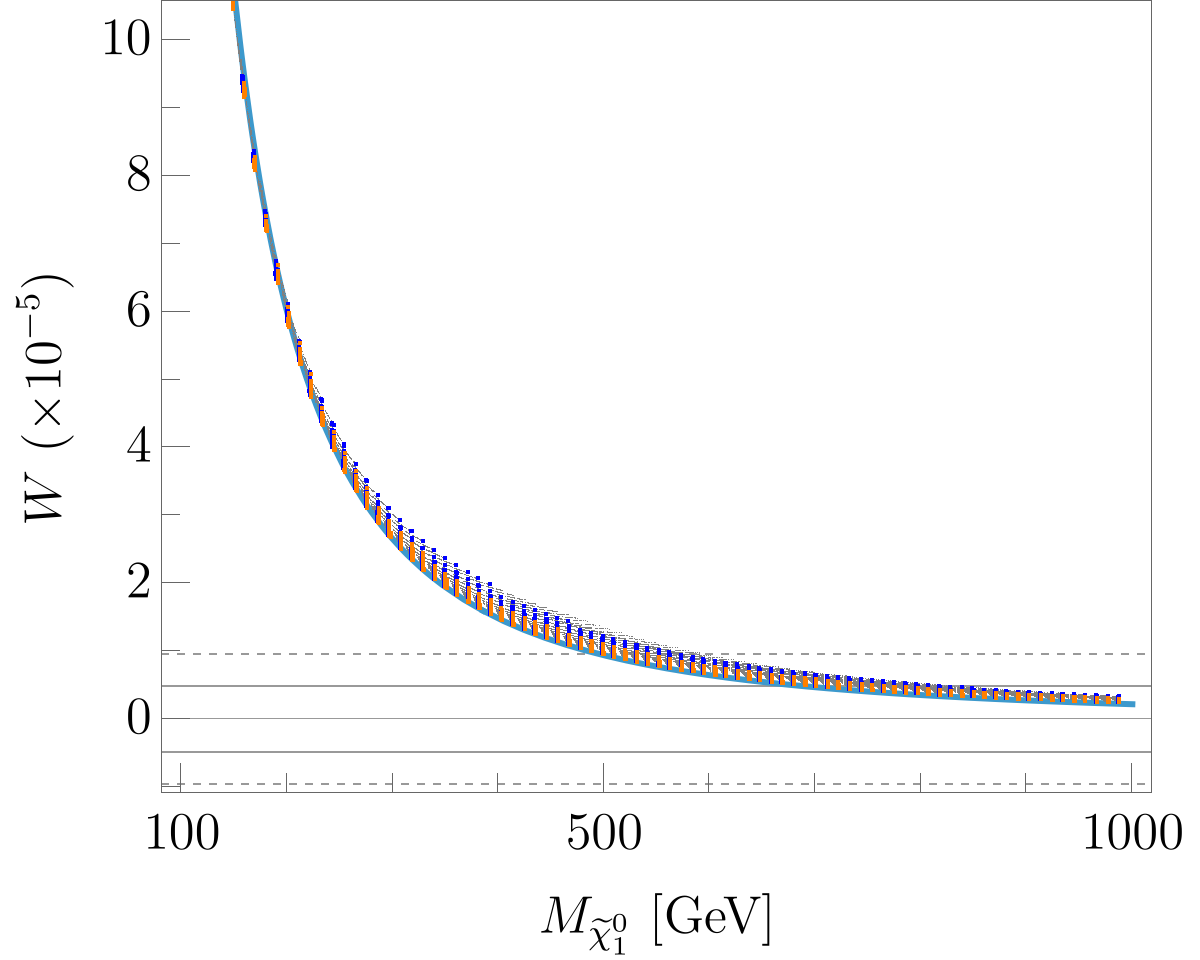}
  \caption{
    The \(W\) parameter as a function of the higgsino-like LSP mass. The color coding of the points is the same as that used in Fig.~\ref{fig:oblique}: gray points are excluded by the current LZ bound, orange points lie within the neutrino fog, and blue points predict direct detection cross sections above the neutrino fog. The horizontal gray solid (dashed) lines indicate the projected FCC-ee \(1\sigma\) (\(2\sigma\)) sensitivity given in Eq.~\eqref{eq:1sigs1}. The blue solid curve, partly obscured by the orange data points, shows the approximate expression in Eq.~\eqref{eq:wyappr}.
} 
  \label{fig:WvmLSP}
\end{figure}

Figure~\ref{fig:WvmLSP} shows the \(W\) parameter as a function of the higgsino-like LSP mass. The color coding of the points is the same as that used in Fig.~\ref{fig:oblique}. The horizontal gray solid (dashed) lines indicate the projected FCC-ee \(1\sigma\) (\(2\sigma\)) sensitivity given in Eq.~\eqref{eq:1sigs1}. The blue solid curve, partly obscured by the orange data points, shows the approximate expression in Eq.~\eqref{eq:wyappr}. We find that higgsino-like states with masses $\lesssim 500~\mathrm{GeV}$ can induce a \(W\) parameter large enough to be probed at the \(2\sigma\) level by future electroweak precision measurements.


\begin{figure}
  \centering
  \subcaptionbox{\label{fig:sinth} $\Delta \sin^2 \theta_{\mathrm{eff}}$ }{
  \includegraphics[width=0.48\columnwidth]{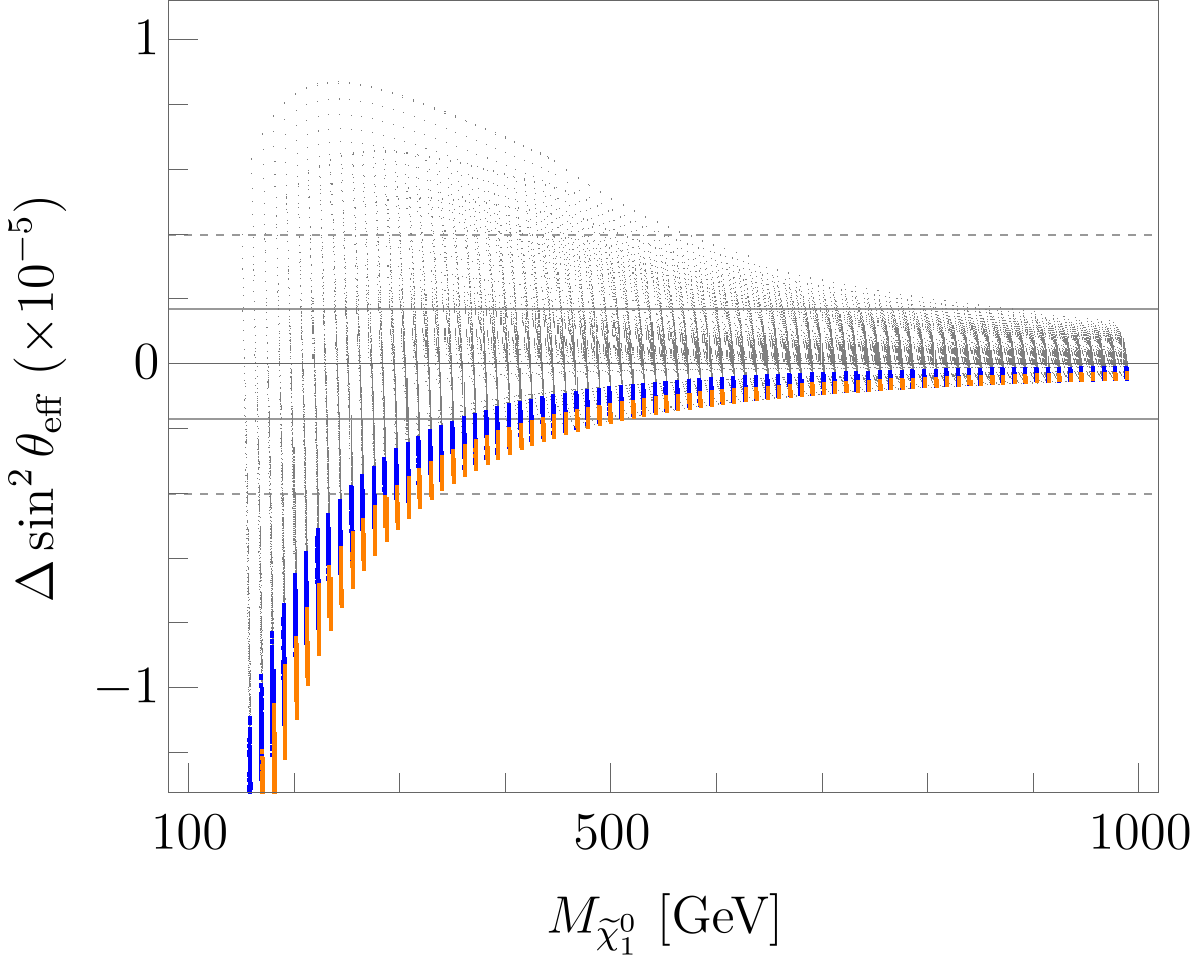}}
  \subcaptionbox{\label{fig:mw} $\Delta M_W$}{
  \includegraphics[width=0.48\columnwidth]{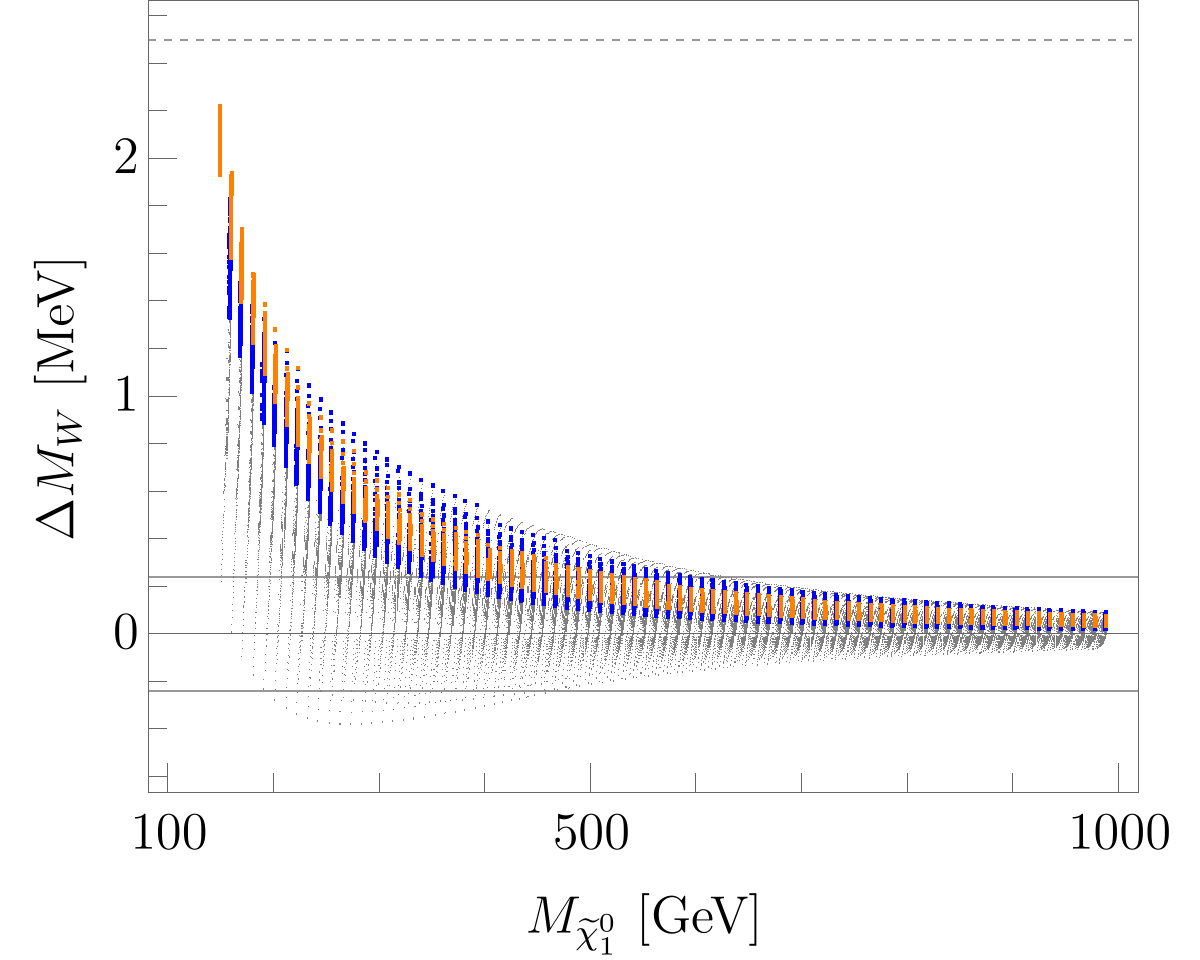}} 
  \\[10pt] 
  \subcaptionbox{\label{fig:mwvssinth} $\Delta M_W$ vs $\Delta \sin^2 \theta_{\mathrm{eff}}$}{
  \includegraphics[width=0.5\columnwidth]{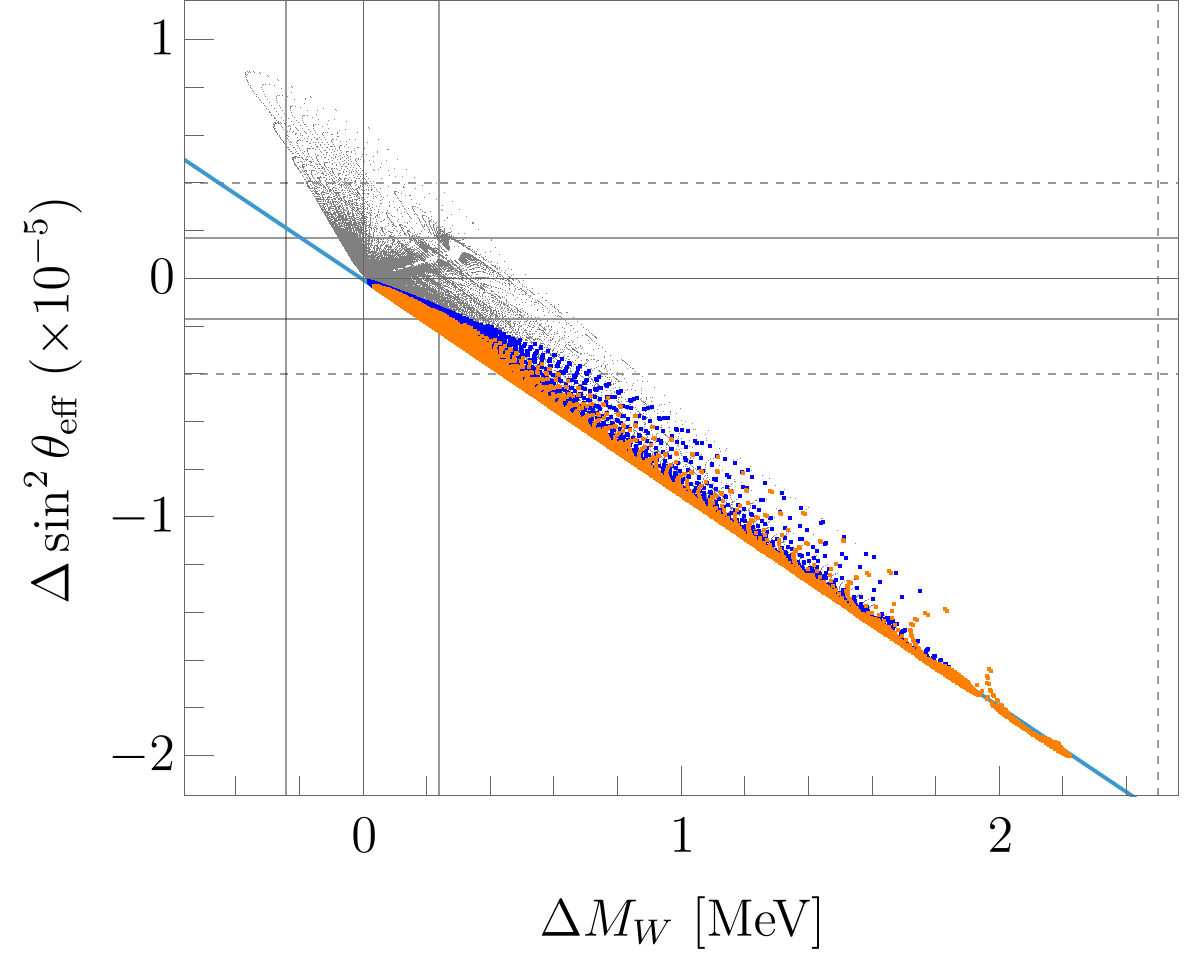}}
\caption{
Predicted shifts in the EWPOs obtained in our parameter scan; 
(a) the shift in the effective weak mixing angle, \(\Delta \sin^2 \theta_{\mathrm{eff}}\), as a function of the higgsino-like LSP mass; (b) the shift in the \(W\)-boson mass, \(\Delta M_W\), as a function of the higgsino-like LSP mass; (c) the correlation between \(\Delta M_W\) and \(\Delta \sin^2 \theta_{\mathrm{eff}}\). The color coding of the points is the same as that used in Fig.~\ref{fig:oblique}: gray points are excluded by the current LZ bound, orange points lie within the neutrino fog, and blue points predict direct detection cross sections above the neutrino fog. The horizontal gray solid (dashed) lines in Fig.~\ref{fig:sinth} and Fig.~\ref{fig:mwvssinth} indicate the projected FCC-ee (Giga-Z) sensitivities. The horizontal (vertical) gray solid and dashed lines in Fig.~\ref{fig:mw} (Fig.~\ref{fig:mwvssinth}) indicate the projected FCC-ee and ILC 250 sensitivities, respectively. The blue line in Fig.~\ref{fig:mwvssinth} corresponds to the relation in Eq.~\eqref{eq:relswmw}.
} 
\label{fig:mwandsinth}
\end{figure}

Figure~\ref{fig:mwandsinth} shows the predicted shifts in the EWPOs, \(\Delta \sin^2 \theta_{\mathrm{eff}}\) and \(\Delta M_W\), obtained in our parameter scan. In Fig.~\ref{fig:sinth}, the horizontal axis represents the higgsino-like LSP mass, while the vertical axis shows the shift in the effective weak mixing angle, \(\Delta \sin^2 \theta_{\mathrm{eff}}\). Similarly, Fig.~\ref{fig:mw} displays the shift in the \(W\)-boson mass, \(\Delta M_W\), as a function of the higgsino-like LSP mass. Figure~\ref{fig:mwvssinth} shows the correlation between \(\Delta M_W\) and \(\Delta \sin^2 \theta_{\mathrm{eff}}\). The color coding of the points is the same as that used in Fig.~\ref{fig:oblique}. The horizontal gray solid (dashed) lines in Figs.~\ref{fig:sinth} and \ref{fig:mwvssinth} indicate the projected FCC-ee (Giga-$Z$) sensitivities. The horizontal (vertical) gray solid and dashed lines in Fig.~\ref{fig:mw} (Fig.~\ref{fig:mwvssinth}) indicate the projected FCC-ee and ILC 250 sensitivities, respectively. The blue line in Fig.~\ref{fig:mwvssinth} corresponds to the relation in Eq.~\eqref{eq:relswmw}. 

We find that higgsino contributions generally predict negative shifts in \(\sin^2 \theta_{\mathrm{eff}}\) and positive shifts in \(M_W\). Although positive values of \(\Delta \sin^2 \theta_{\mathrm{eff}}\) and negative values of \(\Delta M_W\) are possible before imposing the dark matter direct detection constraints, such parameter points are excluded by the current LZ bound. Consequently, after applying the direct detection constraints, the allowed parameter space leads to rather characteristic deviations from the SM predictions. We also find that the predicted shifts in \(\sin^2 \theta_{\mathrm{eff}}\) can exceed not only the projected FCC-ee sensitivity but also that of the Giga-$Z$ program. At the same time, the figures illustrate that observing deviations of this magnitude will require substantial reductions in the current theoretical uncertainties given in Eq.~\eqref{eq:therrornow}. This highlights the importance of future progress in higher-order calculations as well as improved determinations of the SM input parameters in fully exploiting the precision reach of future $e^+ e^-$ colliders.

Figure~\ref{fig:mwvssinth} further demonstrates that the predicted shifts in \(\Delta M_W\) and \(\Delta \sin^2 \theta_{\mathrm{eff}}\) exhibit a strong correlation. This behavior originates from the fact that the dominant contributions in the present scenario arise mainly through the \(W\) and \(Y\) parameters. Indeed, by setting $\hat{S} = \hat{T} =  0$ in Eqs.~\eqref{eq:delmw} and \eqref{eq:delsw} and using Eq.~\eqref{eq:relyw}, we obtain
\begin{equation}
  \Delta \sin^2 \theta_{\mathrm{eff}} \simeq - 2 \cos^2 \theta_W \cdot \sin^2 \theta_{\mathrm{eff}} \cdot \frac{\Delta M_W^2}{M_W^2} ~.
  \label{eq:relswmw}
\end{equation}
This relation leads to a striking and characteristic prediction of the higgsino scenario considered in this work.

\begin{figure}
  \centering
  \subcaptionbox{  
  \label{fig:sicross_neg} $\mu<0$ }{
  \includegraphics[width=0.45\columnwidth]{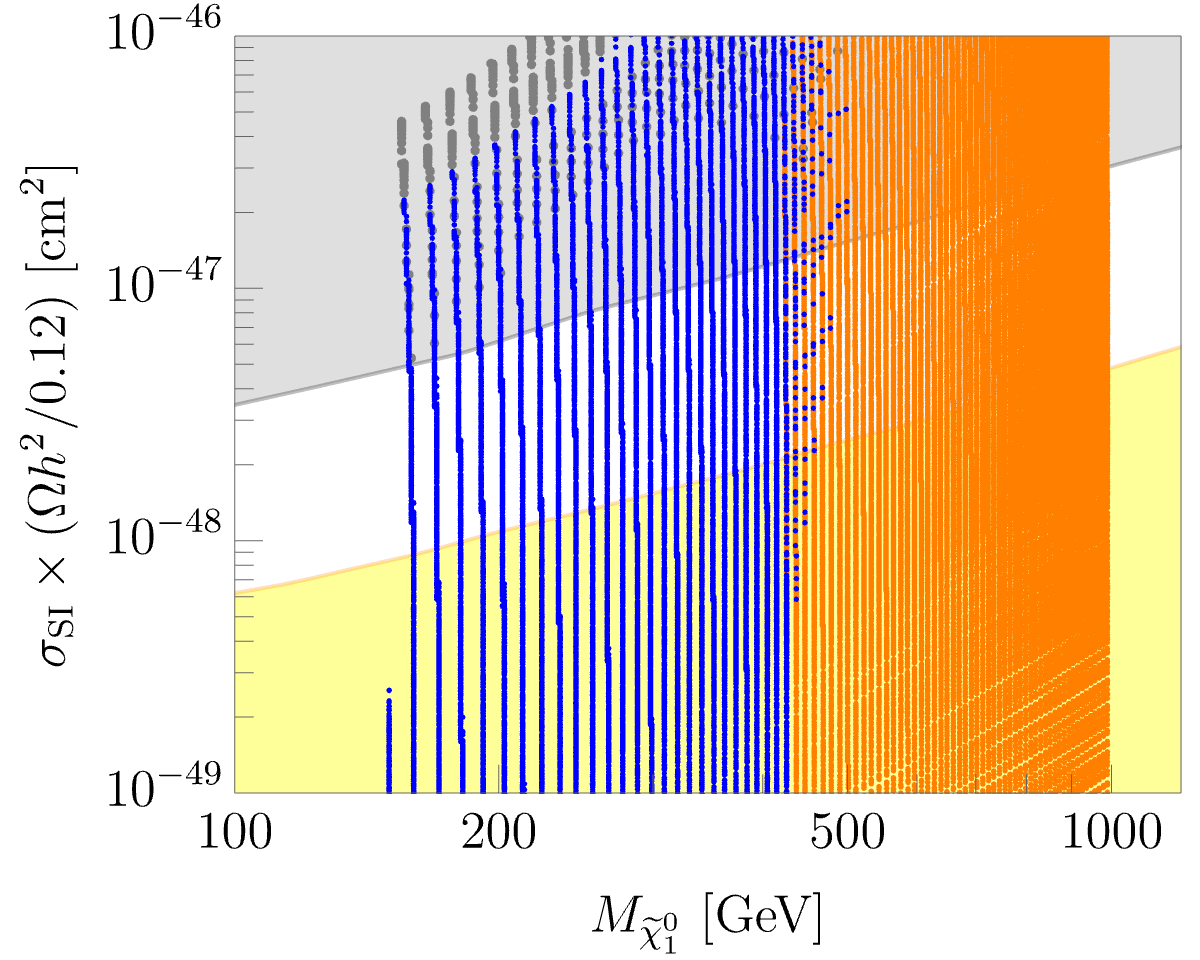}}
  \subcaptionbox{  
  \label{fig:sicross_pos} $\mu>0$}{
  \includegraphics[width=0.45\columnwidth]{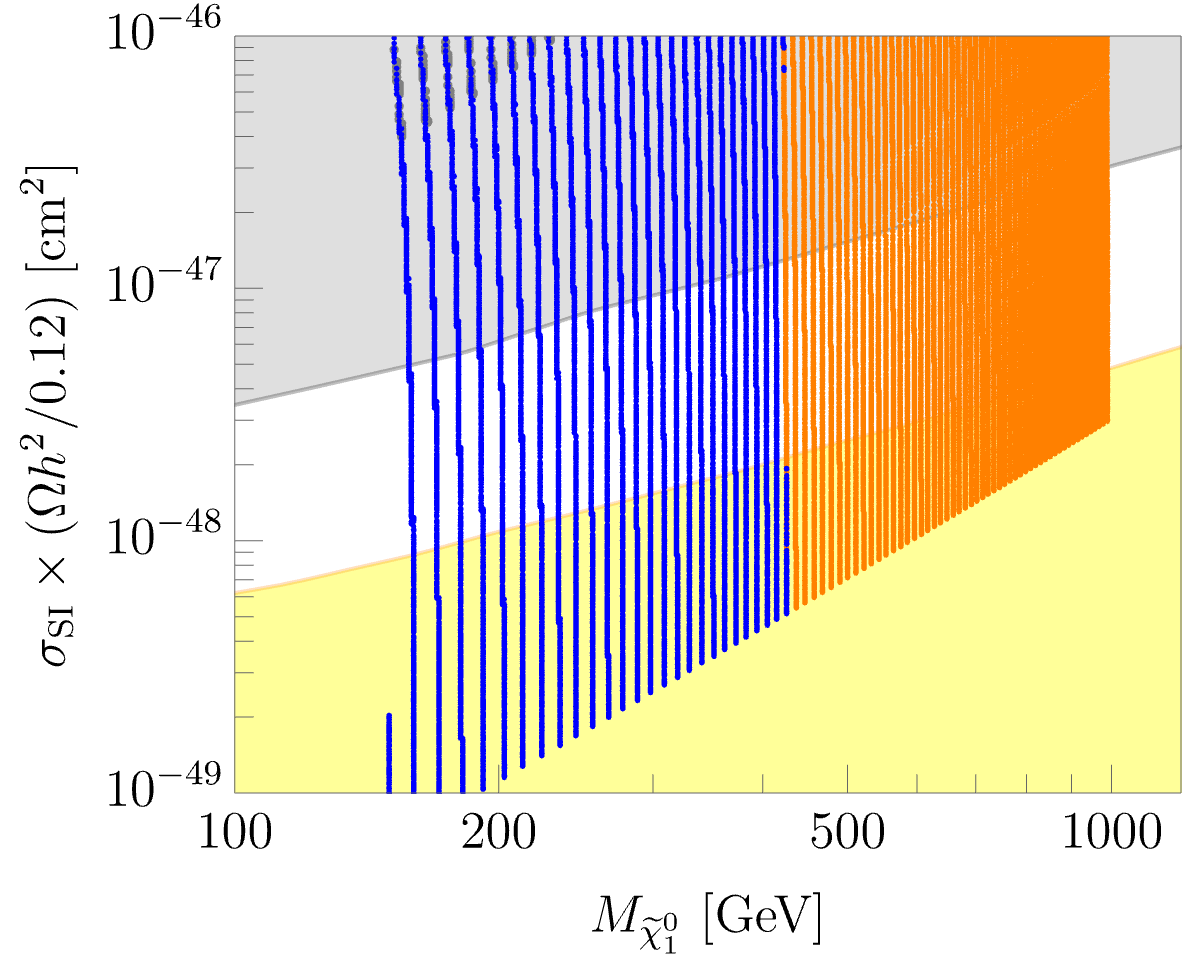}}
  \caption{
    SI dark matter-nucleon scattering cross section as a function of the dark matter mass for (a) \(\mu < 0 \) and (b) \(\mu > 0\). The gray points are excluded by the current LZ limit on the SD scattering cross section. The blue points correspond to parameter regions that predict deviations larger than \(2\sigma\) in future electroweak precision measurements, while the orange points correspond to parameter regions with smaller deviations. The gray shaded region indicates the current LZ exclusion bound, whereas the yellow shaded region represents the neutrino fog.
  }
  \label{fig:sicross}
\end{figure}

Figure~\ref{fig:sicross} shows the scatter plot of the SI dark matter-nucleon scattering cross section as a function of the higgsino-like dark matter mass for \(\mu < 0\) (Fig.~\ref{fig:sicross_neg}) and \(\mu > 0\) (Fig.~\ref{fig:sicross_pos}). The gray points are excluded by the current LZ limit on the SD scattering cross section. The blue points correspond to parameter regions that predict deviations larger than \(2\sigma\) in future electroweak precision measurements, while the orange points correspond to parameter regions with smaller deviations. In evaluating the sensitivity of electroweak precision measurements, we use the \(\chi^2\) function defined in Eq.~\eqref{eq:chi2} for the S1 scenario. The gray shaded region indicates the current LZ exclusion bound, whereas the yellow shaded region represents the neutrino fog. From these plots, we clearly see that electroweak precision measurements and dark matter direct detection experiments play complementary roles in probing higgsinos. In particular, future electroweak precision measurements retain sensitivity to parameter regions in which the direct detection cross section is strongly suppressed and approaches the neutrino fog. This demonstrates that EWPOs can probe higgsino scenarios that are difficult to access through direct detection experiments alone.

\begin{figure}
  \centering
  \subcaptionbox{  
  \label{fig:delm_neg} $\mu<0$ }{
  \includegraphics[width=0.45\columnwidth]{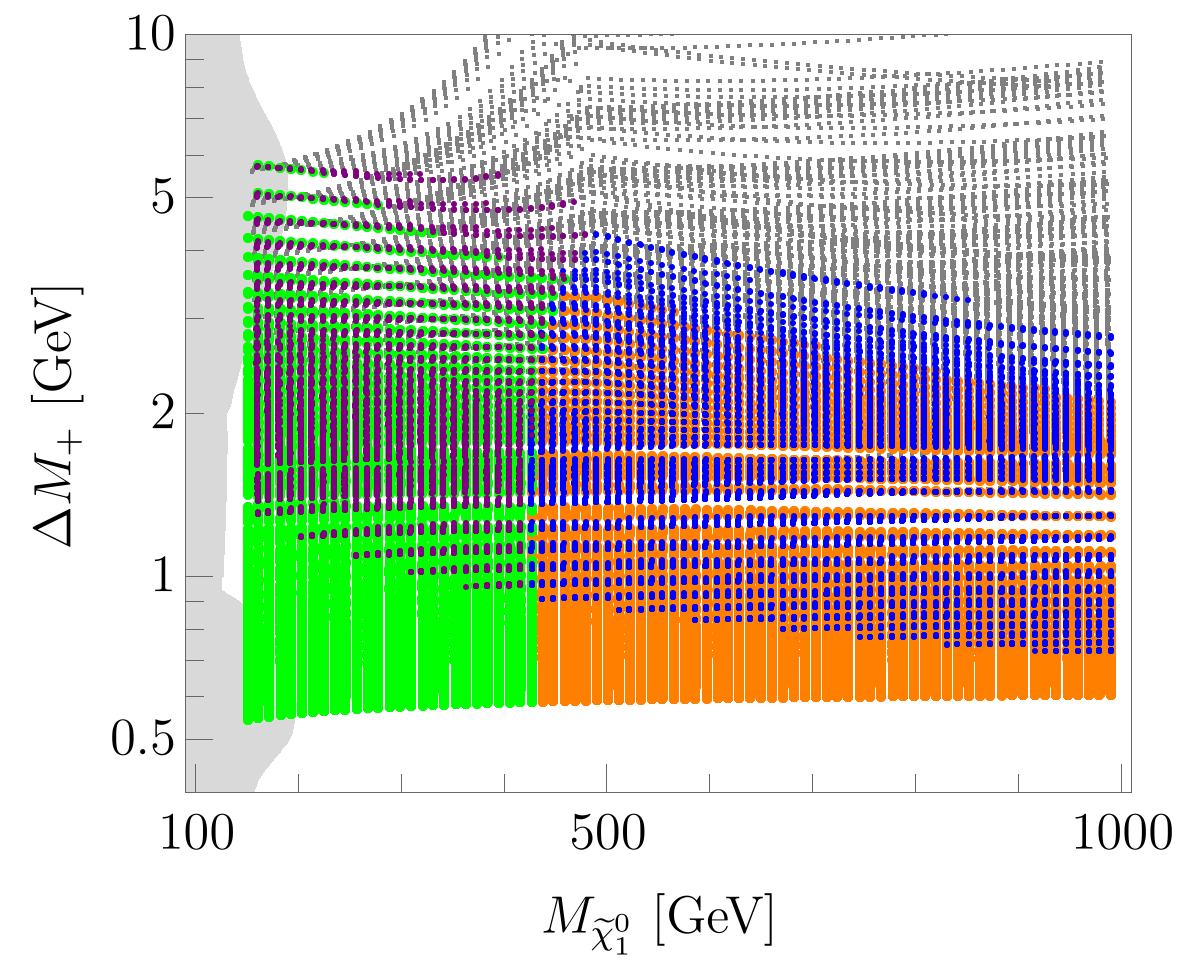}}
  \subcaptionbox{  
  \label{fig:delm_pos} $\mu>0$}{
  \includegraphics[width=0.45\columnwidth]{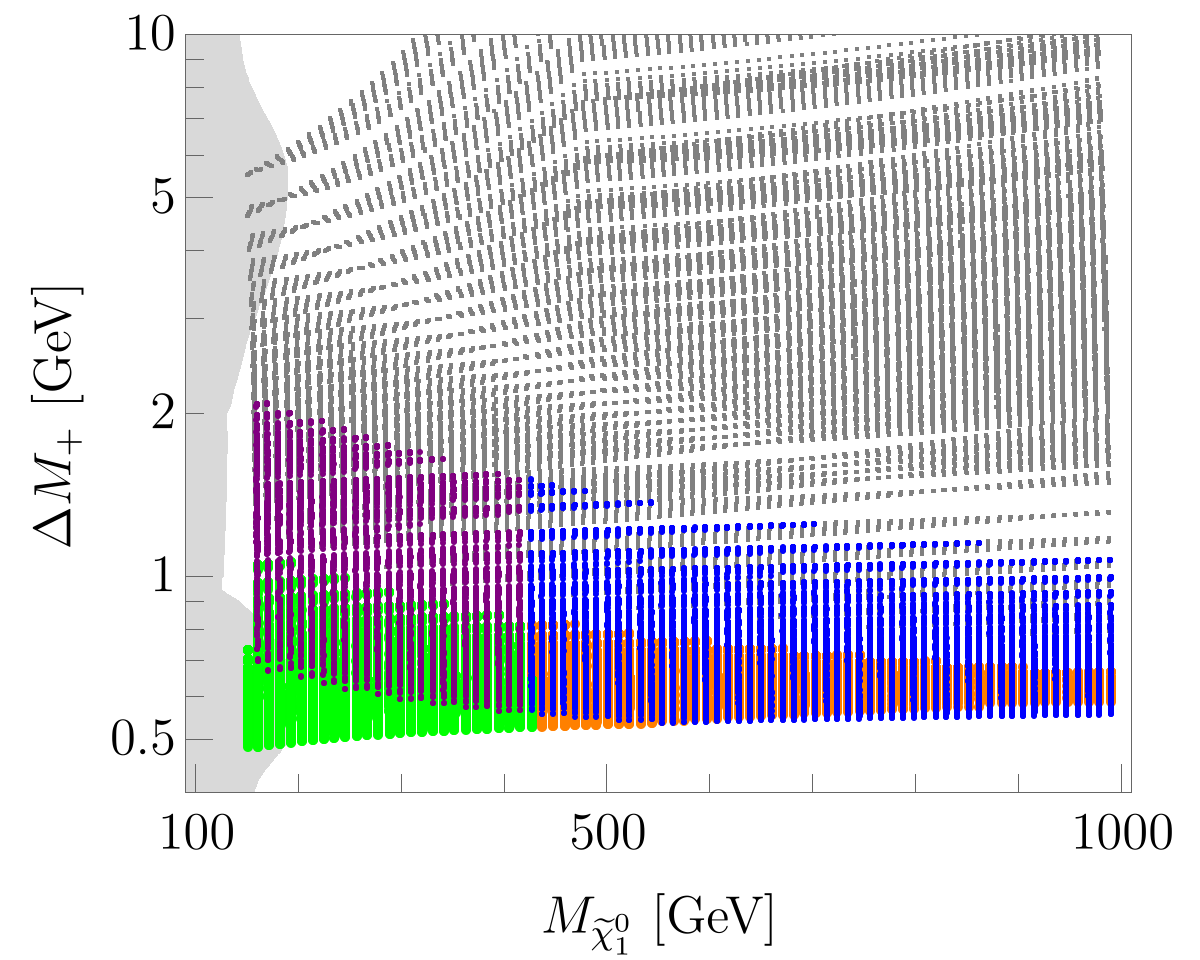}}
  \caption{
Charged-neutral mass splitting as a function of the higgsino-like LSP mass for (a) \(\mu < 0 \) and (b) \(\mu > 0\). Gray points are excluded by the current LZ bound. Blue points correspond to parameter regions that can be probed by future dark matter direct detection experiments but not by future electroweak precision measurements at the \(2\sigma\) level. Green points correspond to parameter regions that are difficult to probe by future direct detection experiments but can be probed by electroweak precision measurements. Purple points correspond to parameter regions that can be probed by both approaches. Orange points correspond to parameter regions that are difficult to probe by both approaches. The gray shaded region indicates the parameter space excluded by current LHC searches~\cite{ATLAS:2025lhc}.
} 
\label{fig:delm}
\end{figure}

Figure~\ref{fig:delm} shows the charged-neutral mass splitting as a function of the higgsino-like LSP mass for \(\mu < 0\) (Fig.~\ref{fig:delm_neg}) and \(\mu > 0\) (Fig.~\ref{fig:delm_pos}). The gray points are excluded by the current LZ bound. The blue points correspond to parameter regions that can be probed by future dark matter direct detection experiments, namely those with SI scattering cross sections above the neutrino fog, but cannot be probed by future electroweak precision measurements in the sense that they induce deviations smaller than \(2\sigma\) in the S1 scenario. On the other hand, the green points correspond to parameter regions that are difficult to probe by future direct detection experiments because their SI scattering cross sections lie below the neutrino fog, while they can still be probed by electroweak precision measurements with deviations larger than \(2\sigma\). The purple points correspond to parameter regions that can be probed by both approaches. The orange points correspond to parameter regions that are difficult to probe by either future dark matter direct detection experiments or electroweak precision measurements. The gray shaded region indicates the parameter space excluded by current LHC searches~\cite{ATLAS:2025lhc}.
We find that, once the current LZ constraints are imposed, the surviving parameter space typically predicts rather small charged-neutral mass splittings. In such compressed spectra, the visible decay products of the chargino become very soft, significantly reducing the sensitivity of conventional LHC searches. Importantly, there remain many parameter points that evade both LHC searches and future dark matter direct detection experiments, but can still induce observable deviations in future electroweak precision measurements. By contrast, dark matter direct detection experiments are most sensitive to scenarios with relatively large charged-neutral mass splittings, corresponding to sizable higgsino-gaugino mixing, and can probe higgsino dark matter up to the thermal relic mass scale, \(M_{\widetilde{\chi}^0_1} \simeq 1~\mathrm{TeV}\). This again highlights the strong complementarity between EW precision measurements and dark matter direct detection experiments in exploring higgsino dark matter scenarios.

\section{Conclusion}
\label{sec:conclusion}

In this paper, we have investigated the prospects for probing light higgsinos through future electroweak precision measurements and dark matter direct detection experiments. We compared the projected sensitivities of future \(e^+e^-\) colliders with those of present and future dark matter direct detection experiments. We found that the dominant sensitivity to higgsinos arises through the \(W\) parameter, while the contributions to \(\hat{S}\) and \(\hat{T}\) are typically suppressed once the current direct detection constraints are imposed. As a result, future electroweak precision measurements can probe higgsinos with masses up to \(\sim 500~\mathrm{GeV}\), including parameter regions in which the SI dark matter-nucleon scattering cross section lies below the neutrino fog. By contrast, dark matter direct detection experiments can efficiently probe scenarios with sizable charged-neutral mass splittings and are capable of testing thermal higgsino dark matter up to \(M_{\widetilde{\chi}^0_1} \simeq 1~\mathrm{TeV}\). Our results therefore highlight the strong complementarity between electroweak precision measurements and dark matter direct detection experiments in exploring light higgsino scenarios. We also showed that the predicted shifts in \(M_W\) and \(\sin^2\theta_{\mathrm{eff}}\) exhibit a strong correlation characteristic of the higgsino scenario considered in this work.

Finally, we emphasized that future improvements in theoretical calculations will play an essential role in fully exploiting the precision reach of future \(e^+e^-\) colliders. In particular, reductions in the theoretical uncertainties of EWPOs will be crucial for establishing sensitivity to the small deviations predicted in light higgsino scenarios.

\section*{Acknowledgments}

The work of N.N. was supported in part by the Grant-in-Aid for Scientific Research C (No. 25K07314).
The work of K.H. was supported in part by JSPS KAKENHI Grant Numbers 24H02244 and 24K07041.
The work of G.O. was supported in part by JSPS KAKENHI Grant Numbers 25KJ0950.

\newpage
\section*{Appendix}
\appendix

\section{Vacuum polarization functions}
\label{app:vacpol}

In this appendix, we summarize the one-loop contributions of neutralinos and charginos to the electroweak gauge boson vacuum polarization functions defined in Eq.~\eqref{eq:vpdef}. The contributions to the transverse parts of the vacuum polarization functions are given by
\begin{align}
  \Pi_{WW} (q^2) &= -\frac{g^2}{16\pi^2} \sum_{i= 1}^{2} \sum_{j=1}^{4} \Bigl[\bigl( \bigl|\mathcal{C}_{ij}^L\bigr|^2 + \bigl|\mathcal{C}_{ij}^R\bigr|^2 \bigr) H ( q^2, M^2_{\widetilde{\chi}^+_i}, M^2_{\widetilde{\chi}^0_j} ) \nonumber \\ 
  &+ 4 \mathrm{Re} \bigl( \mathcal{C}_{ij}^L \mathcal{C}_{ij}^{R*}  \bigr) M_{\widetilde{\chi}^+_i} M_{\widetilde{\chi}^0_j}B_0 ( q^2, M^2_{\widetilde{\chi}^+_i}, M^2_{\widetilde{\chi}^0_j} )
  \Bigr] ~, \\ 
  \Pi_{ZZ} (q^2) &= -\frac{g^2}{16\pi^2 \cos^2 \theta_W}  \biggl[\sum_{i,j=1}^{2} \Bigl\{
      ( \bigl|\mathcal{G}_{ij}^L\bigr|^2 + \bigl|\mathcal{G}_{ij}^R\bigr|^2 \bigr) H ( q^2, M^2_{\widetilde{\chi}^+_i}, M^2_{\widetilde{\chi}^+_j} ) \nonumber \\ 
      &+ 4 \mathrm{Re} \bigl( \mathcal{G}_{ij}^L \mathcal{G}_{ij}^{R*}  \bigr) M_{\widetilde{\chi}^+_i} M_{\widetilde{\chi}^+_j}B_0 ( q^2, M^2_{\widetilde{\chi}^+_i}, M^2_{\widetilde{\chi}^+_j} ) \Bigr\}\nonumber \\ 
      &+\sum_{i,j=1}^{4} \Bigl\{ \bigl|\mathcal{H}_{ij}^L\bigr|^2  H ( q^2, M^2_{\widetilde{\chi}^0_i}, M^2_{\widetilde{\chi}^0_j} )
      - 2\mathrm{Re} \bigl[\bigl( \mathcal{H}_{ij}^L  \bigr)^2\bigr] M_{\widetilde{\chi}^0_i} M_{\widetilde{\chi}^0_j}B_0 ( q^2, M^2_{\widetilde{\chi}^0_i}, M^2_{\widetilde{\chi}^0_j} ) \Bigr\}
  \biggr] ~, \\ 
  \Pi_{Z\gamma} (q^2) &= -\frac{eg}{16\pi^2 \cos \theta_W} \sum_{i=1}^{2} \bigl(\mathcal{G}_{ii}^L + \mathcal{G}_{ii}^R\bigr) \Bigl[H ( q^2, M^2_{\widetilde{\chi}^+_i}, M^2_{\widetilde{\chi}^+_i} ) + 2 M_{\widetilde{\chi}^+_i}^2 B_0 ( q^2, M^2_{\widetilde{\chi}^+_i}, M^2_{\widetilde{\chi}^+_i} ) \Bigr] ~, 
  \\ 
  \Pi_{\gamma\gamma} (q^2) &= -\frac{e^2}{8\pi^2 } \sum_{i=1}^{2}  \Bigl[H ( q^2, M^2_{\widetilde{\chi}^+_i}, M^2_{\widetilde{\chi}^+_i} ) + 2 M_{\widetilde{\chi}^+_i}^2 B_0 ( q^2, M^2_{\widetilde{\chi}^+_i}, M^2_{\widetilde{\chi}^+_i} ) \Bigr] ~.
\end{align}
The couplings appearing above are defined as
\begin{align}
  \mathcal{C}_{ij}^L&=-V_{i1}N_{j2}^*+\frac{1}{\sqrt{2}}V_{i2}N^*_{j4}~,
 \nonumber \\ 
  \mathcal{C}_{ij}^R&=-U_{i1}^*N_{j2}-\frac{1}{\sqrt{2}}U_{i2}^*N_{j3}~, \nonumber \\ 
  \mathcal{G}^L_{ij}&=
  V_{i1}V_{j1}^* +\frac{1}{2}V_{i2}V^*_{j2}-\delta_{ij}\sin^2\theta_W~,
  \nonumber \\
   \mathcal{G}^R_{ij}&=
  U_{i1}^*U_{j1} +\frac{1}{2}U_{i2}^*U_{j2}-\delta_{ij}\sin^2\theta_W~,
  \nonumber \\
  \mathcal{H}^L_{ij}&=
  \frac{1}{2}(N_{i3}N_{j3}^*-N_{i4}N_{j4}^*)~,
\end{align}
where $e$ denotes the electric charge. The loop function $B_0$ is defined as
\begin{equation}
  B_0 (q^2, m_1^2, m_2^2) \equiv -\int_0^1 dx \, \ln \biggl[\frac{x\, m_1^2 + (1-x) m_2^2 - x(1-x) q^2 - i\epsilon}{Q^2}\biggr] ~,
\end{equation}
where $Q$ denotes the renormalization scale,\footnote{We take $Q = M_Z$ throughout this work.} and $\epsilon$ is a positive infinitesimal quantity. The function $H(q^2,m_1^2,m_2^2)$ is given by
\begin{align}
  H &(q^2, m_1^2, m_2^2) \equiv -\frac{1}{3} \biggl[
      m_1^2 \biggl( \frac{m_1^2-m_2^2}{q^2} - 2\biggr) \ln \biggl(\frac{m_1^2}{Q^2}\biggr)+ m_2^2 \biggl( -\frac{m_1^2-m_2^2}{q^2} - 2\biggr) \ln \biggl(\frac{m_2^2}{Q^2}\biggr) 
   \nonumber \\ 
  &-  \biggl\{
      2 q^2 - (m_1^2 + m_2^2) - \frac{(m_1^2 - m_2^2)^2}{q^2}    
      \biggr\}    B_0 (q^2, m_1^2, m_2^2)+ \frac{2q^2}{3} - \frac{(m_1^2 - m_2^2)^2}{q^2}\biggr] ~.
\end{align}
Some useful formulae for the loop functions $B_0 (q^2, m_1^2, m_2^2)$ and $H (q^2, m_1^2, m_2^2)$, as well as their derivatives, are summarized in the appendix of Ref.~\cite{Nagata:2025ycf}. We also note the following relations:
\begin{align}
  \Pi_{Z\gamma} (0) &= \Pi_{\gamma\gamma}(0) = 0 ~, \\
  \Pi'_{\gamma\gamma} (0) &= \frac{e^2}{8\pi^2} \cdot \biggl(\frac{2}{3}\biggr) \sum_i \ln \biggl(\frac{M^2_{\widetilde{\chi}^+_i}}{Q^2}\biggr) ~, \\ 
  \Pi'_{Z\gamma} (0) &= \frac{e g}{16\pi^2 \cos\theta_W }  \cdot \biggl(\frac{2}{3}\biggr) \sum_{i=1}^{2} \bigl(\mathcal{G}_{ii}^L + \mathcal{G}_{ii}^R\bigr)\ln \biggl(\frac{M^2_{\widetilde{\chi}^+_i}}{Q^2}\biggr) ~.
\end{align}

\section{Non-thermal dark matter scenario}
\label{app:nonthermal}

In this appendix, we consider a non-thermal dark matter scenario in which the observed dark matter abundance is supplemented by non-thermally produced higgsinos, rather than being determined solely by the thermal relic abundance. In this case, the suppression factor associated with the thermal relic density in Eq.~\eqref{eq:sigma_eff} is absent, and consequently the constraints from dark matter direct detection experiments become significantly stronger.

\begin{figure}
  \centering
  \subcaptionbox{\label{fig:sinth_nt} $\Delta \sin^2 \theta_{\mathrm{eff}}$ }{
  \includegraphics[width=0.48\columnwidth]{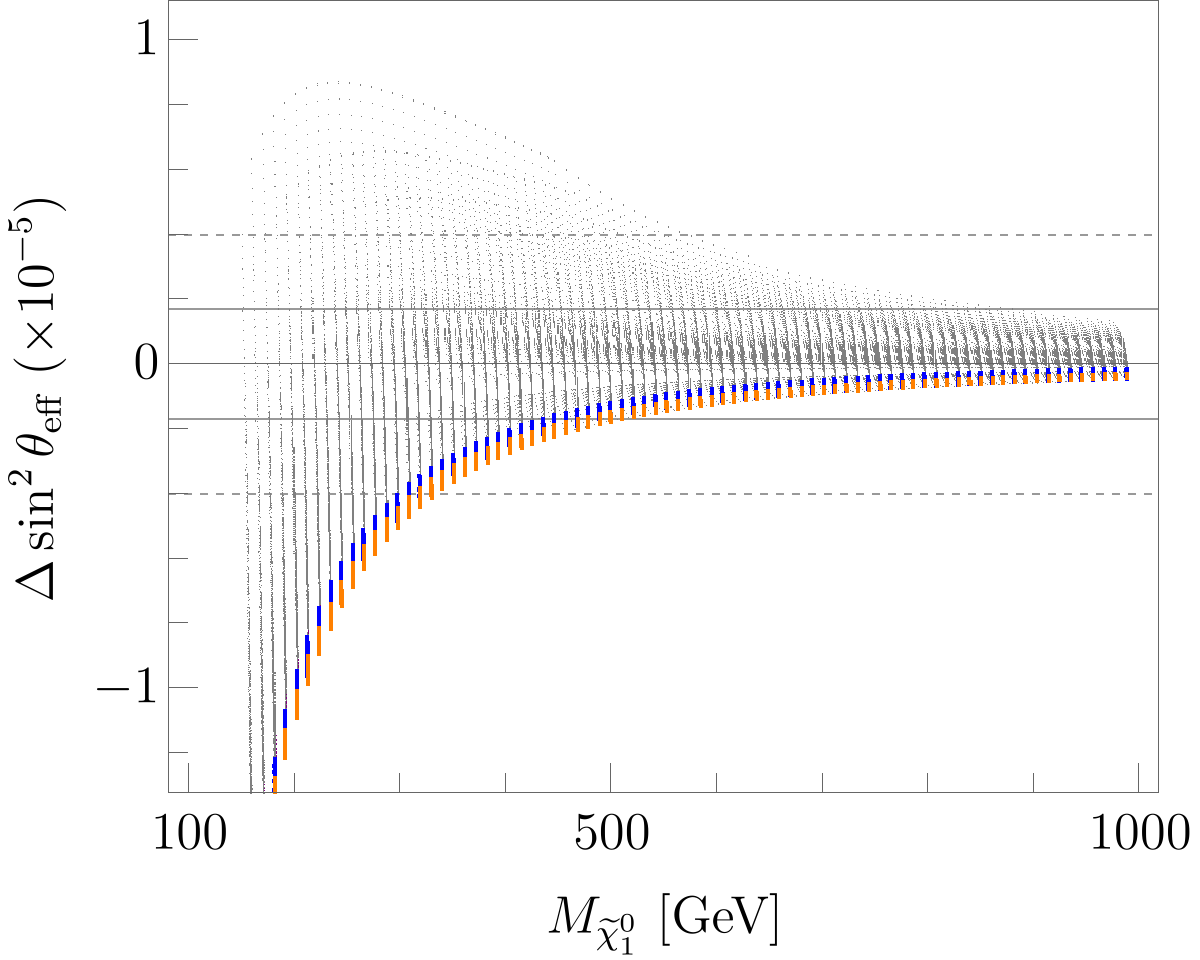}}
  \subcaptionbox{\label{fig:mw_nt} $\Delta M_W$}{
  \includegraphics[width=0.48\columnwidth]{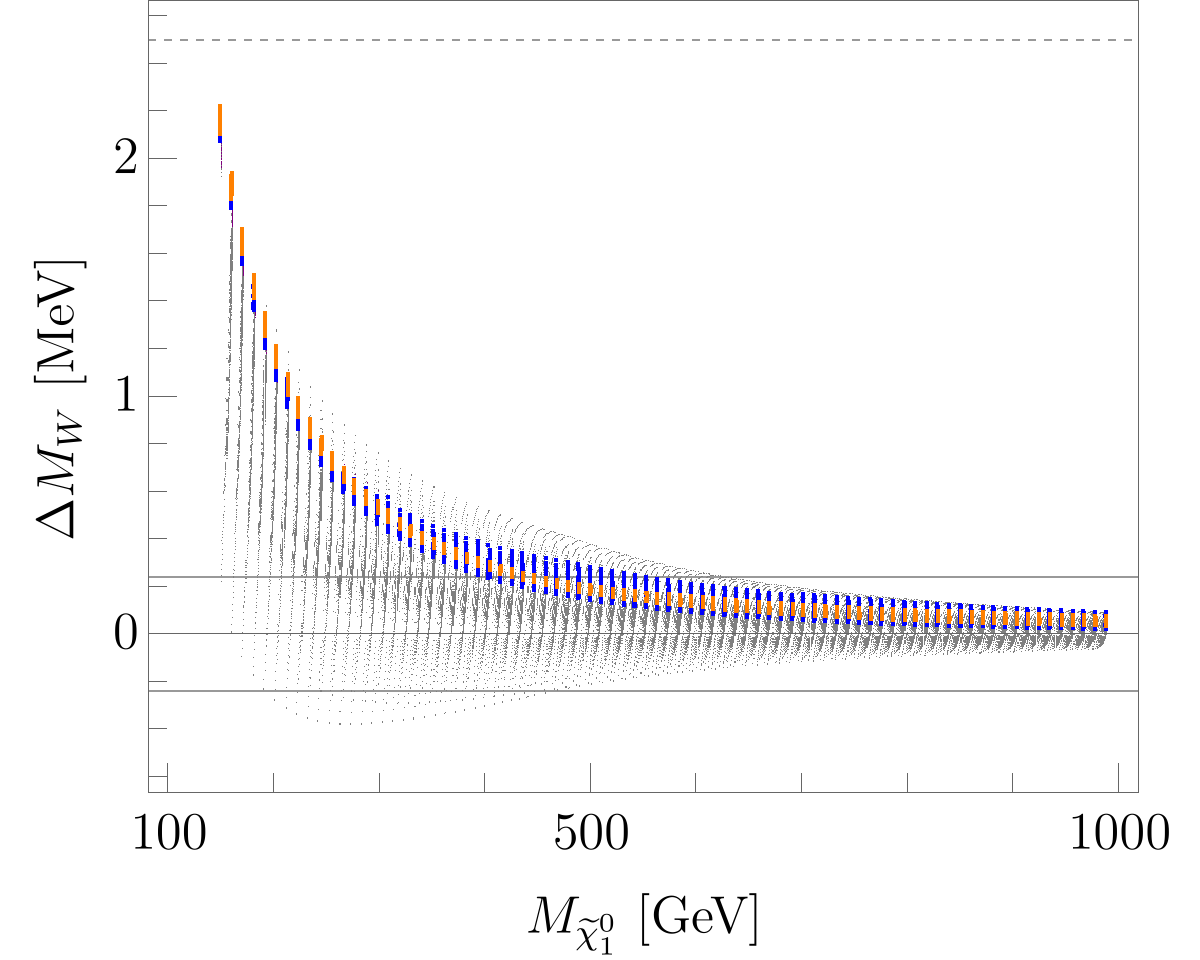}} 
\caption{
Predicted shifts in the EWPOs obtained in our parameter scan for the non-thermal dark matter scenario; 
(a) the shift in the effective weak mixing angle, \(\Delta \sin^2 \theta_{\mathrm{eff}}\), as a function of the dark matter mass; (b) the shift in the \(W\)-boson mass, \(\Delta M_W\), as a function of the dark matter mass. The color coding of the points is the same as that used in Fig.~\ref{fig:oblique}: gray points are excluded by the current LZ bound, orange points lie within the neutrino fog, and blue points predict direct detection cross sections above the neutrino fog. The gray solid lines indicate the projected FCC-ee sensitivities, whereas the gray dashed lines show the Giga-Z (ILC 250) sensitivity in Fig.~\ref{fig:sinth_nt} (Fig.~\ref{fig:mw_nt}). 
} 
\label{fig:mwandsinth_nt}
\end{figure}

\begin{figure}
  \centering\subcaptionbox{  
  \label{fig:sicross_neg_nt} $\mu<0$ }{
  \includegraphics[width=0.45\columnwidth]{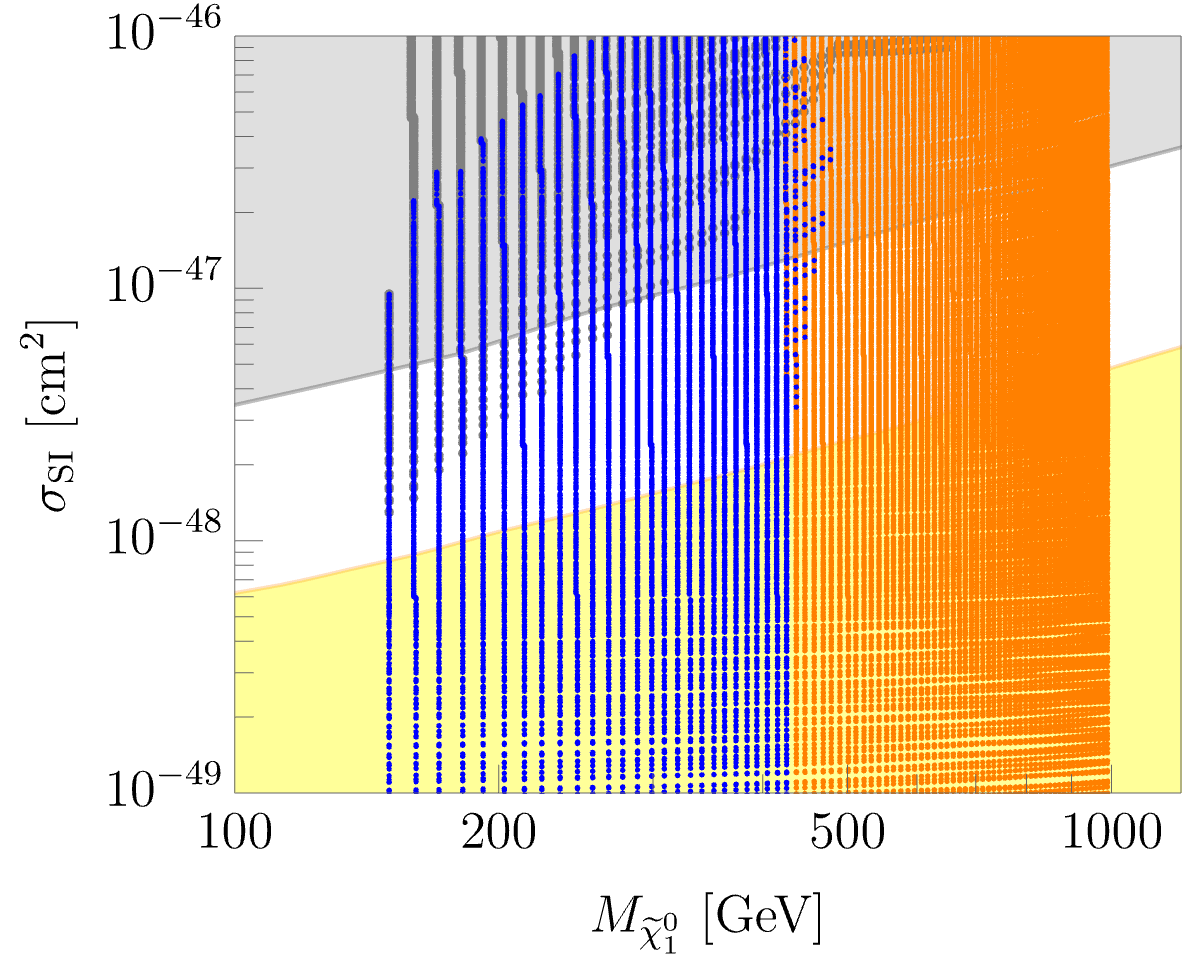}}
  \subcaptionbox{  
  \label{fig:sicross_pos_nt} $\mu>0$}{
  \includegraphics[width=0.45\columnwidth]{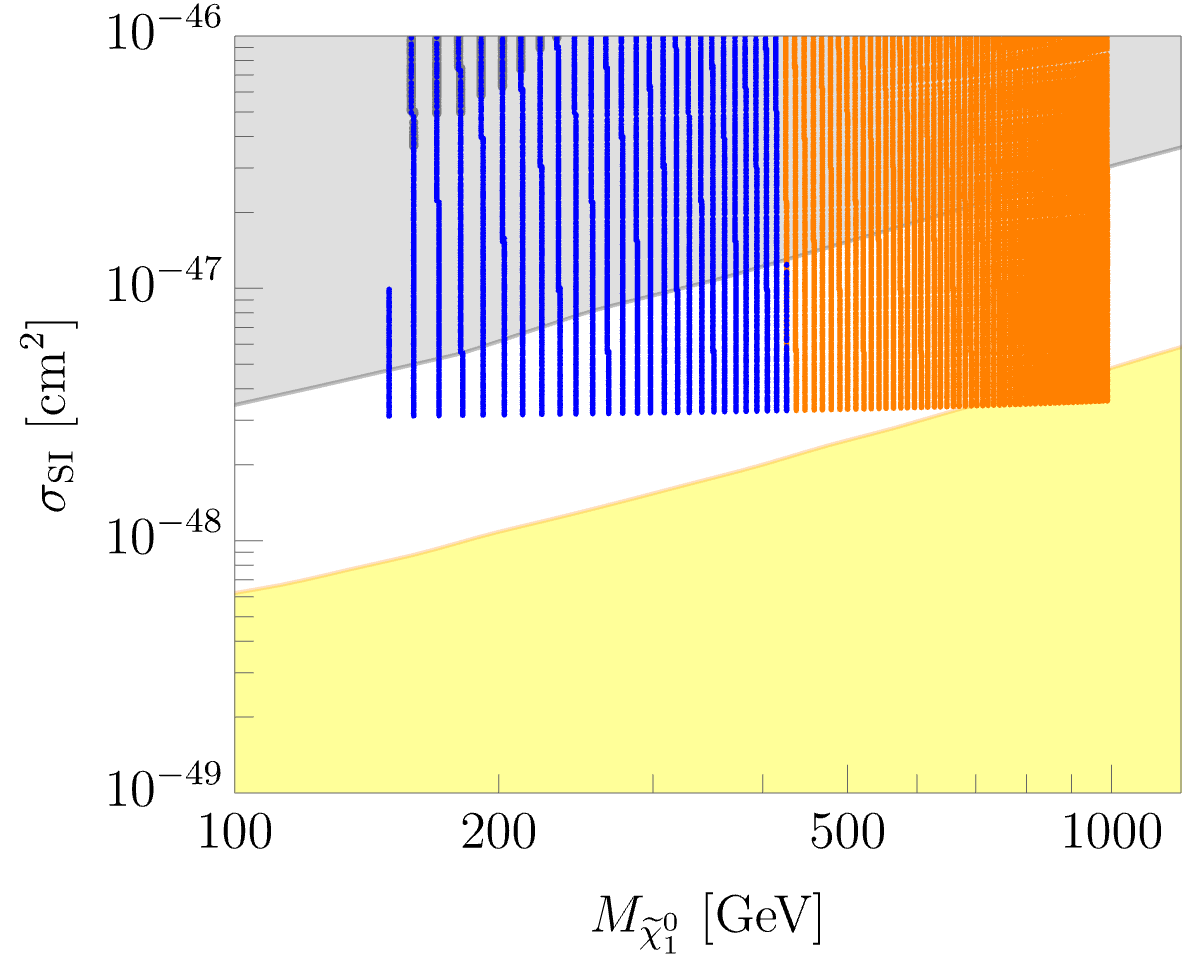}}
  \caption{
    SI dark matter-nucleon scattering cross section as a function of the dark matter mass for the non-thermal dark matter scenario in the case of (a) \(\mu < 0 \) and (b) \(\mu > 0\). The gray points are excluded by the current LZ limit on the SD scattering cross section. The blue points correspond to parameter regions that predict deviations larger than \(2\sigma\) in future electroweak precision measurements, while the orange points correspond to parameter regions with smaller deviations. The gray shaded region indicates the current LZ exclusion bound, whereas the yellow shaded region represents the neutrino fog.
  }  
  \label{fig:sicross_nt}
\end{figure}

\begin{figure}
  \centering
  \subcaptionbox{  
  \label{fig:delm_neg_nt} $\mu<0$ }{
  \includegraphics[width=0.45\columnwidth]{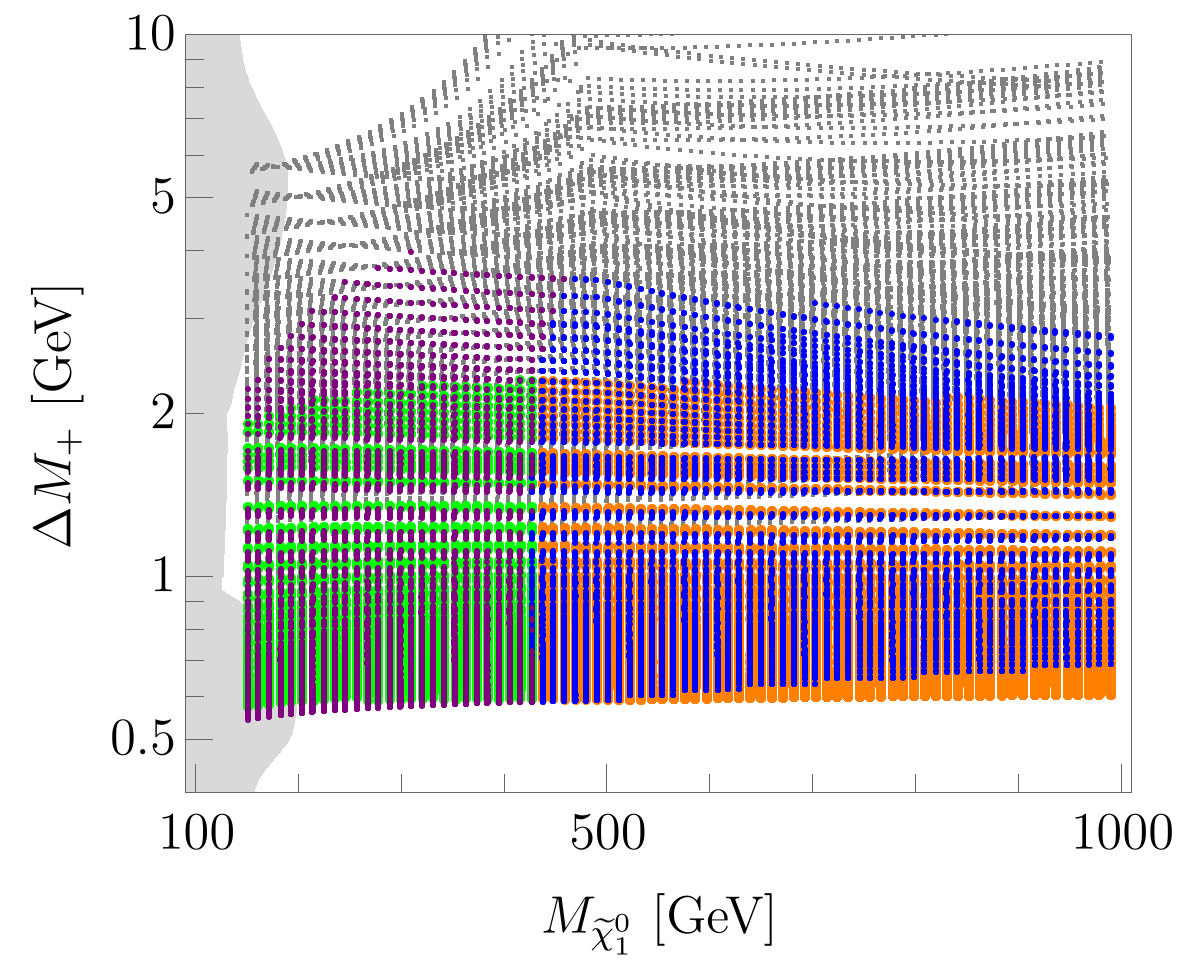}}
  \subcaptionbox{  
  \label{fig:delm_pos_nt} $\mu>0$}{
  \includegraphics[width=0.45\columnwidth]{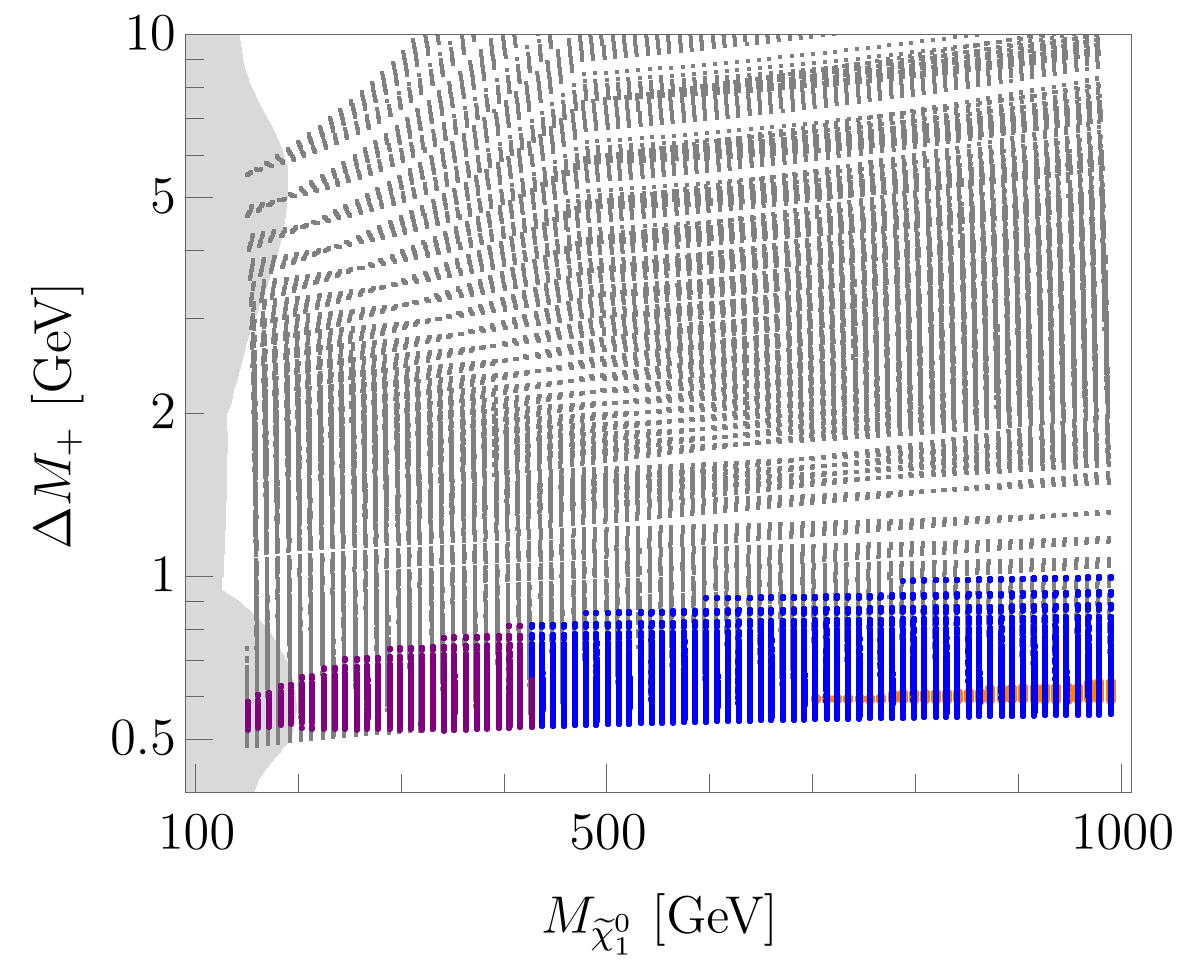}}
  \caption{
Charged-neutral mass splitting as a function of the dark matter mass for the non-thermal dark matter scenario in the case of (a) \(\mu < 0 \) and (b) \(\mu > 0\). Gray points are excluded by the current LZ bound. Blue points correspond to parameter regions that can be probed by future dark matter direct detection experiments but not by future electroweak precision measurements at the \(2\sigma\) level. Green points correspond to parameter regions that are difficult to probe by future direct detection experiments but can be probed by electroweak precision measurements. Orange points correspond to parameter regions that are difficult to probe by both approaches. The gray shaded region indicates the parameter space excluded by current LHC searches.
} 
  \label{fig:delm_nt}
\end{figure}

Figure~\ref{fig:mwandsinth_nt} shows the non-thermal counterpart of Fig.~\ref{fig:mwandsinth}. Figures~\ref{fig:sinth_nt} and \ref{fig:mw_nt} display the predicted shifts in \(\Delta \sin^2 \theta_{\mathrm{eff}}\) and \(\Delta M_W\), respectively.
Figure~\ref{fig:sicross_nt} shows the non-thermal counterpart of Fig.~\ref{fig:sicross}, while Fig.~\ref{fig:delm_nt} shows the non-thermal counterpart of Fig.~\ref{fig:delm}. The color coding of the points in these figures is identical to that used in the corresponding figures in Sec.~\ref{sec:results}. 
The lower edge of the scattered points in Fig.~\ref{fig:sicross_pos_nt} simply reflects the upper limit of the scan range, $M_1 < 10~\mathrm{TeV}$, and does not correspond to a physical boundary.
It is worth noting that some parameter points in Fig.~\ref{fig:sicross_neg_nt} that evade the current LZ bound on the SI scattering cross section are nevertheless excluded by the bound on the SD scattering cross section (represented by the gray points). A similar observation was reported in Ref.~\cite{Martin:2024ytt}.

Compared with the thermal scenario discussed in Sec.~\ref{sec:results}, the constraints from the current LZ experiment become significantly stronger in the non-thermal case. This is because the direct detection event rate is no longer suppressed by the small thermal relic abundance of higgsino dark matter, and the full local dark matter density is assumed to consist of higgsinos. Consequently, a much larger fraction of the parameter space is excluded by the present direct detection limits, particularly in regions with sizable higgsino-gaugino mixing that enhance the Higgs-mediated scattering cross section.

\section{Alternative Uncertainty Scenarios}
\label{app:theor_unc}

In this appendix, we consider several benchmark scenarios for future theoretical and systematic uncertainties proposed in Ref.~\cite{Greljo:2025ggc}, beyond the S1 scenario adopted in Sec.~\ref{sec:uncertainty}. By comparing the resulting sensitivities to light higgsinos, we assess the impact of future improvements in theoretical calculations and experimental systematics on the reach of electroweak precision measurements.

The S2 scenario of Ref.~\cite{Greljo:2025ggc} assumes the same perturbative theory uncertainties as the S1 scenario, while neglecting the uncertainties associated with the conversion of experimental measurements into EWPOs. It therefore corresponds to a situation in which the extraction of EWPOs from experimental data is no longer a limiting factor and the overall precision is determined primarily by the uncertainties of the SM theoretical predictions. In this case, the projected uncertainties and correlation matrix are given by
\begin{align}
  \begin{pmatrix}
    \sigma_{\hat{S}} & \sigma_{\hat{T}} & \sigma_W & \sigma_Y
  \end{pmatrix}
  = 
  \begin{pmatrix}
    1.74 & 0.73 & 0.47 & 1.55
  \end{pmatrix}
  \times 10^{-5} ~,
\end{align}
and
\begin{align}
  \rho = 
  \begin{pmatrix}
    1 & 0.825 & 0.394 & 0.937 \\ 
    0.825 & 1 & 0.248 & 0.63 \\ 
    0.394 & 0.248 & 1 & 0.216 \\ 
    0.937 & 0.63 & 0.216 & 1 
  \end{pmatrix}
  ~,
\end{align}
respectively.

The S3 scenario of Ref.~\cite{Greljo:2025ggc} represents the idealized limit in which both perturbative theory uncertainties and uncertainties associated with the extraction of EWPOs are negligible, corresponding to the ultimate experimental sensitivity. The projected uncertainties and correlation matrix in this scenario are given by
\begin{align}
  \begin{pmatrix}
    \sigma_{\hat{S}} & \sigma_{\hat{T}} & \sigma_W & \sigma_Y
  \end{pmatrix}
  = 
  \begin{pmatrix}
    1.71 & 0.63 & 0.46 & 1.55
  \end{pmatrix}
  \times 10^{-5} ~,
\end{align}
and
\begin{align}
  \rho = 
  \begin{pmatrix}
    1 & 0.885 & 0.414 & 0.952 \\ 
    0.885 & 1 & 0.365 & 0.749 \\ 
    0.414 & 0.365 & 1 & 0.211 \\ 
    0.952 & 0.749 & 0.211 & 1 
  \end{pmatrix}
  ~,
\end{align}
respectively.

For comparison, we also consider the projected electroweak fit presented in Ref.~\cite{Maura:2024zxz}, which we refer to as the S4 scenario throughout this appendix. The corresponding projected uncertainties and correlation matrix are given by
\begin{align}
  \begin{pmatrix}
    \sigma_{\hat{S}} & \sigma_{\hat{T}} & \sigma_W & \sigma_Y
  \end{pmatrix}
  = 
  \begin{pmatrix}
    2.26 & 1.16 & 0.49 & 1.90
  \end{pmatrix}
  \times 10^{-5} ~,
\end{align}
and
\begin{align}
  \rho = 
  \begin{pmatrix}
    1 & 0.795 & 0.331 & 0.868 \\ 
    0.795 & 1 & 0.088 & 0.461 \\ 
    0.331 & 0.088 & 1 & 0.256 \\ 
    0.868 & 0.461 & 0.256 & 1 
  \end{pmatrix}
  ~,
\end{align}
respectively.

\begin{figure}
  \centering
  \includegraphics[width=0.5\columnwidth]{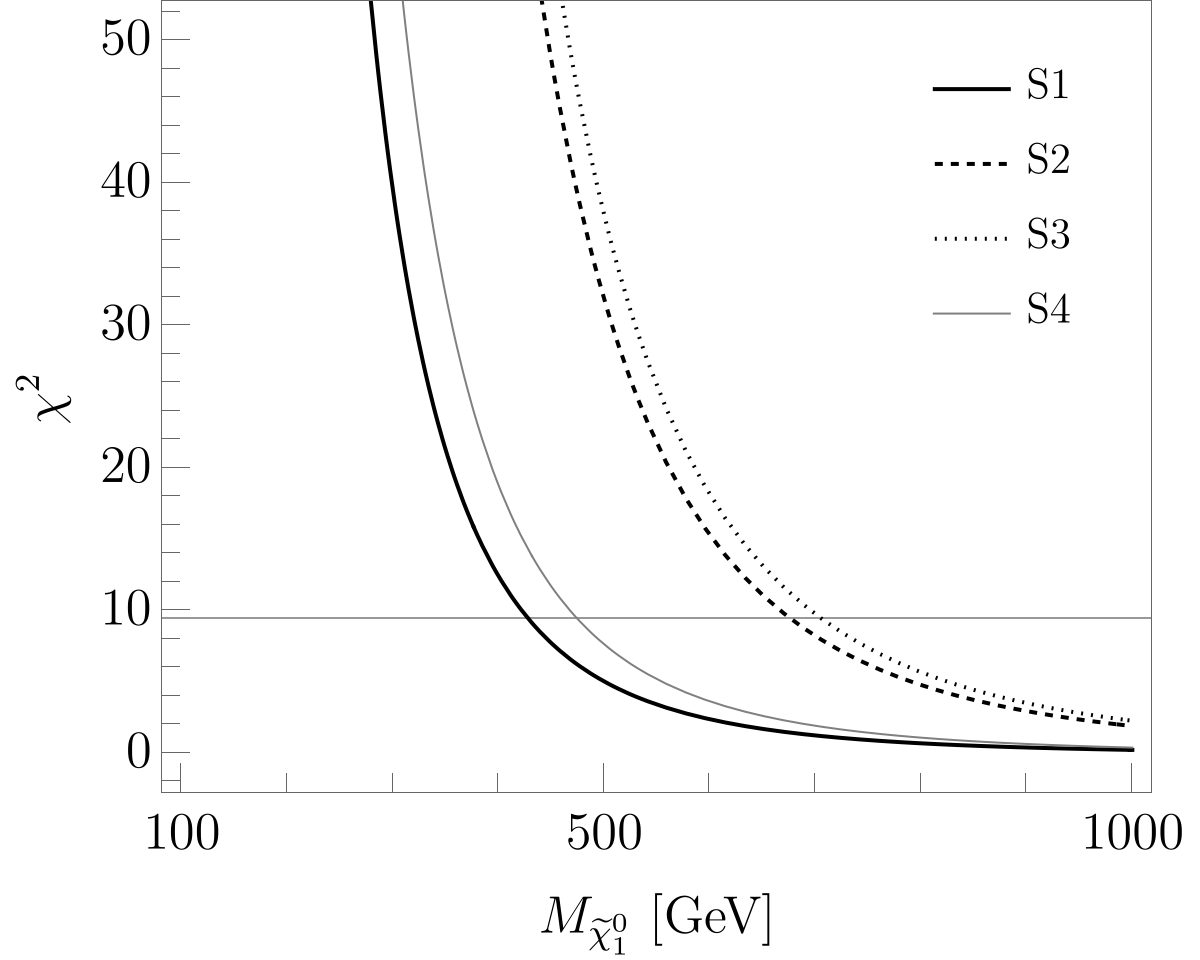}
  \caption{
  The \(\chi^2\) value as a function of the higgsino-like LSP mass for the various benchmark scenarios considered in this paper, assuming the pure-higgsino limit. The horizontal solid line indicates the \(2\sigma\) exclusion threshold.
} 
  \label{fig:mLSPvChipure}
\end{figure}

Figure~\ref{fig:mLSPvChipure} shows the \(\chi^2\) value as a function of the higgsino-like LSP mass for the benchmark scenarios discussed above, assuming the pure-higgsino limit. In this limit, the oblique parameters are given by Eq.~\eqref{eq:wyappr} with \(\hat{S}=\hat{T}=0\). The horizontal solid line indicates the \(2\sigma\) exclusion threshold. As can be seen from the figure, the S2 and S3 scenarios yield a substantially greater mass reach than the S1 scenario, extending the sensitivity to higgsinos by up to several hundred GeV. This demonstrates the important role played by future advances in theoretical calculations and uncertainty control in determining the ultimate sensitivity of electroweak precision measurements to higgsino-induced effects. We also find that the mass reach obtained in the S4 scenario is in good agreement with that reported in Ref.~\cite{Maura:2024zxz},  providing a useful cross-check of our analysis.

\begin{figure}
  \centering
  \subcaptionbox{\label{fig:mLSPvChiS1} S1 }{
  \includegraphics[width=0.48\columnwidth]{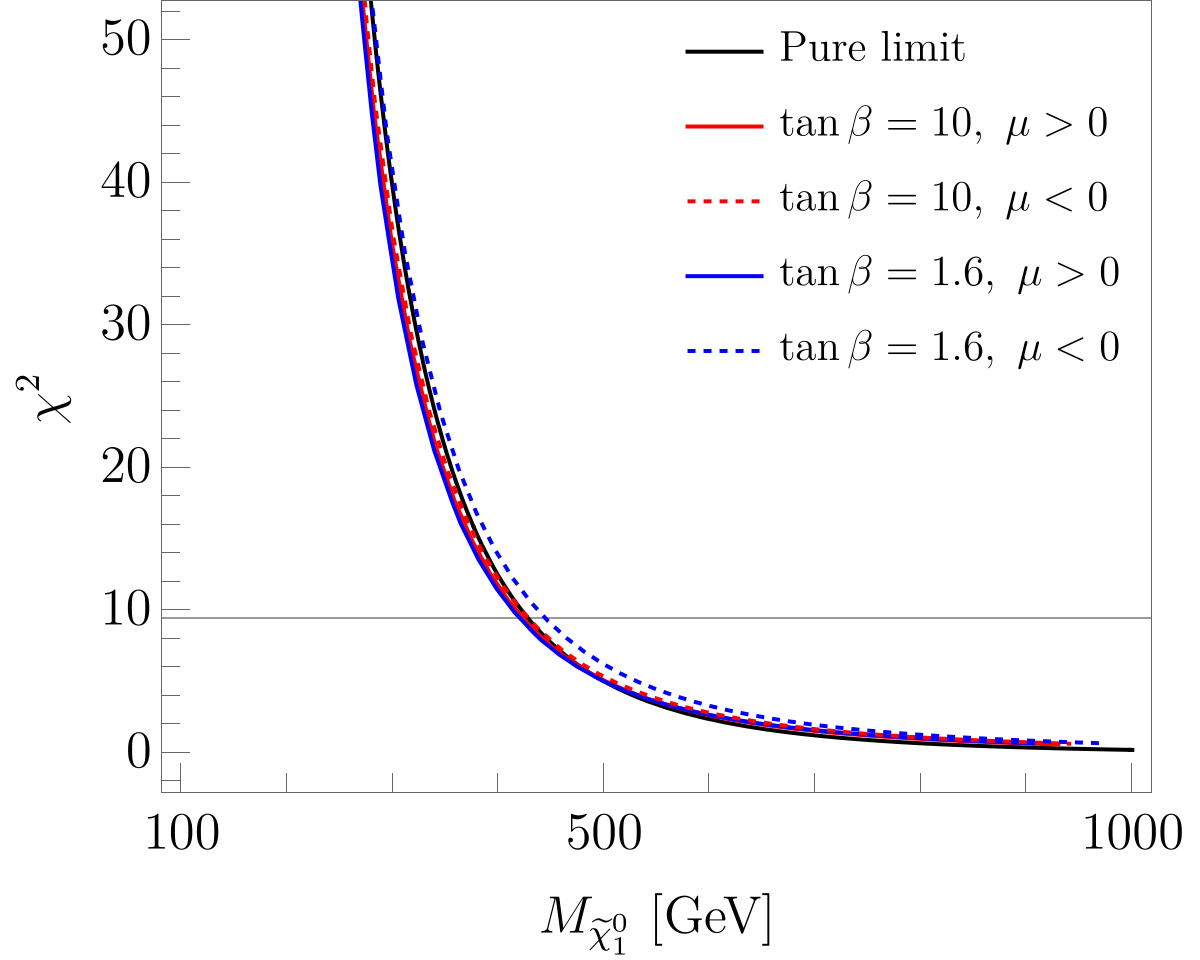}}
  \subcaptionbox{\label{fig:mLSPvChiS3} S3}{
  \includegraphics[width=0.48\columnwidth]{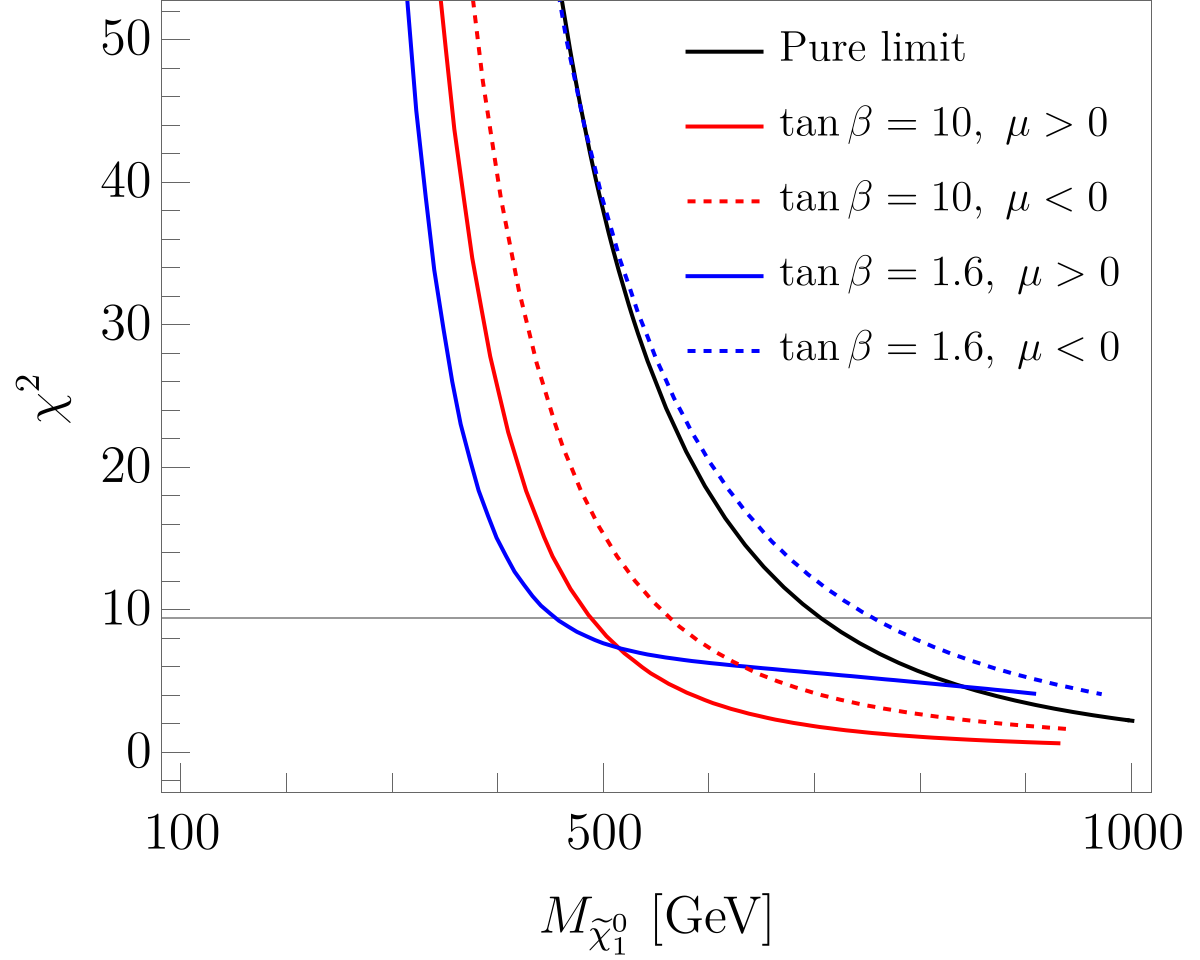}} 
\caption{
  \(\chi^2\) as a function of the higgsino-like LSP mass for the S1 (left) and S3 (right) scenarios. The red (blue) curves correspond to \(\tan\beta=10\) (\(1.6\)). Solid (dashed) lines denote positive (negative) values of \(\mu\). The black solid curve represents the pure-higgsino limit. We fix \(M_1=1~\mathrm{TeV}\) and \(M_2=2~\mathrm{TeV}\).
} 
\label{fig:mLSPvChi}
\end{figure}

Figure~\ref{fig:mLSPvChi} shows the \(\chi^2\) value as a function of the higgsino-like LSP mass for the S1 (Fig.~\ref{fig:mLSPvChiS1}) and S3 (Fig.~\ref{fig:mLSPvChiS3}) scenarios. The red and blue curves correspond to \(\tan\beta=10\) and \(1.6\), respectively. Solid (dashed) lines denote positive (negative) values of \(\mu\). The black solid curve represents the pure-higgsino limit. Throughout this analysis, we fix \(M_1=1~\mathrm{TeV}\) and \(M_2=2~\mathrm{TeV}\). From Fig.~\ref{fig:mLSPvChiS1}, we see that the projected sensitivity is largely insensitive to the choice of \(\tan\beta\) and the sign of \(\mu\). In contrast, the sensitivity in the S3 scenario exhibits a more noticeable dependence on \(\tan\beta\) and the sign of \(\mu\). For \(\tan\beta=10\), our result is consistent with that reported in Ref.~\cite{Greljo:2025ggc}. We also find that the case with \(\tan\beta=1.6\) and \(\mu<0\) yields a sensitivity that is remarkably close to the pure-higgsino limit. This can be understood from the fact that the Higgs coupling to the higgsino-like LSP is strongly suppressed in this region of parameter space, leading to suppressed contributions to \(\hat{S}\) and \(\hat{T}\). Consequently, the electroweak precision observables are dominated by the \(W\) and \(Y\) parameters, and the resultant sensitivity closely resembles that of a pure higgsino.

\newpage
\bibliographystyle{utphysmod}
\bibliography{ref}


\end{document}